\documentclass[fleqn,usenatbib]{mnras}

\usepackage[T1]{fontenc}
\usepackage{ae,aecompl}

\DeclareRobustCommand{\VAN}[3]{#2}
\let\VANthebibliography\thebibliography
\def\thebibliography{\DeclareRobustCommand{\VAN}[3]{##3}\VANthebibliography}

\usepackage{graphicx}	
\usepackage{amsmath}	
\usepackage{amssymb}	

\usepackage{subfig}
\usepackage{xcolor}

\usepackage{scalerel}
\usepackage{tikz}
\usetikzlibrary{svg.path}

\defcitealias{1997ApJ...484..560J}{JS97}

\definecolor{orcidlogocol}{HTML}{A6CE39}
\tikzset{
        orcidlogo/.pic={
                \fill[orcidlogocol] svg{M256,128c0,70.7-57.3,128-128,128C57.3,256,0,198.7,0,128C0,57.3,57.3,0,128,0C198.7,0,256,57.3,256,128z};
                \fill[white] svg{M86.3,186.2H70.9V79.1h15.4v48.4V186.2z}
                svg{M108.9,79.1h41.6c39.6,0,57,28.3,57,53.6c0,27.5-21.5,53.6-56.8,53.6h-41.8V79.1z M124.3,172.4h24.5c34.9,0,42.9-26.5,42.9-39.7c0-21.5-13.7-39.7-43.7-39.7h-23.7V172.4z}
                svg{M88.7,56.8c0,5.5-4.5,10.1-10.1,10.1c-5.6,0-10.1-4.6-10.1-10.1c0-5.6,4.5-10.1,10.1-10.1C84.2,46.7,88.7,51.3,88.7,56.8z};
        }
}

\newcommand\orcidicon[1]{\href{https://orcid.org/#1}{\mbox{\scalerel*{
                                \begin{tikzpicture}[yscale=-1,transform shape]
                                \pic{orcidlogo};
                                \end{tikzpicture}
                        }{1}}}}

\usepackage{newtxtext,newtxmath}

\title[Weak lensing and the Hubble tension]{Cosmology from weak lensing alone and implications for the Hubble tension}

\author[Alex Hall]{
Alex Hall\thanks{E-mail: ahall@roe.ac.uk} \orcidicon{0000-0002-3139-8651}
\\
Institute for Astronomy, University of Edinburgh, Royal Observatory, Blackford Hill, Edinburgh, EH9 3HJ, UK
}

\date{Accepted XXX. Received YYY; in original form ZZZ}

\pubyear{2021}

\begin{document}
\label{firstpage}
\pagerange{\pageref{firstpage}--\pageref{lastpage}}
\maketitle

\begin{abstract}
We investigate the origin of $\Lambda$CDM parameter constraints in weak lensing, with a focus on the Hubble constant. We explain why current cosmic shear data are sensitive to the parameter combination $S_8 \propto \sigma_8 \Omega_m^{0.5}$, improving upon previous studies through use of the halo model. Motivated by the ongoing discrepancy in measurements of the Hubble constant from high and low redshift, we explain why cosmic shear provides almost no constraint on $H_0$ by showing how the lensing angular power spectrum depends on physical length scales in the dark matter distribution. We derive parameter constraints from galaxy lensing in KiDS and cosmic microwave background weak lensing from Planck and SPTpol, separately and jointly, showing how degeneracies between $\sigma_8$ and $\Omega_m$ can be broken. Using lensing and Big Bang Nucleosynthesis to calibrate the sound horizon measured in projection by baryon acoustic oscillations gives $H_0 = 67.4 \pm 0.9 \; \mathrm{km} \, \mathrm{s}^{-1} \, \mathrm{Mpc}^{-1}$, consistent with previous results from Planck and the Dark Energy Survey. We find that a toy Euclid-like lensing survey provides only weak constraints on the Hubble constant due to degeneracies with other parameters that affect the shape of the lensing correlation functions. If external priors on $n_s$, the baryon density, and the amplitude of baryon feedback are available then sub-percent $H_0$ constraints are achievable with forthcoming lensing surveys.
%
%
\end{abstract}




\section{Introduction}
\label{sec:intro}

Cosmic shear is a powerful tool for constraining cosmological models via the geometry and mass distribution of the Universe~\citep[see][for reviews]{2001PhR...340..291B, 2008PhR...462...67M, 2010CQGra..27w3001B, 2015RPPh...78h6901K}. Current galaxy surveys with lensing-quality imaging and photometric redshifts can place useful constraints on certain combinations of $\Lambda$CDM parameters, as well as on simple extensions to the standard cosmological model~\citep{2018PhRvD..98d3526A, 2018PhRvD..98d3528T, 2019PhRvD..99l3505A, 2019PASJ...71...43H, 2020PASJ...72...16H, 2021A&A...646A.140H, 2020arXiv201016416T, 2021A&A...645A.104A}. Considerable effort is currently going into forthcoming lensing surveys that aim to place percent-level constraints on dark energy models and neutrino mass~\citep{2011arXiv1110.3193L, 2020A&A...642A.191E, 2019JCAP...02..056A}.

For years now the `gold standard' for constraining cosmological models has been the anisotropies in the cosmic microwave background (CMB), supplemented by low-redshift probes of the background expansion rate. Despite this, extracting information on the post-recombination Universe from the CMB is fundamentally limited by its nature as a two-dimensional projected field. While there is a wealth of information in secondary effects such as the thermal and kinetic Sunyaev-Zeldovich effects, CMB lensing, secondary scattering from reionization, and other non-linear sources~\citep[e.g.][]{2008RPPh...71f6902A}, all of which are being actively targeted by forthcoming experiments~\citep{2014JLTP..176..733M, 2016arXiv161002743A, 2019BAAS...51g...6S}, the relative importance of low-redshift probes for constraining cosmological models is likely to increase substantially in the near future.

Weak lensing is an attractive probe of large-scale structure because of its sensitivity to the \emph{total} mass content of the Universe, in contrast with probes that rely on baryonic tracers such as galaxy clustering or the Lyman-$\alpha$ forest. Modelling the distribution of tracers is complicated even on large quasi-linear scales due to the non-linear aspects of galaxy bias~\citep[e.g.][]{2018PhR...733....1D, 2018ARA&A..56..435W} and redshift-space distortions~\citep{2010PhRvD..82f3522T, 2011MNRAS.417.1913R, 2013PhRvD..87h3509T, 2014arXiv1409.1225S}. While considerable progress has been made in mitigating these uncertainties, having a direct probe of the underlying matter density field is clearly of great value. A caveat to this is that weak lensing measures the \emph{projected} density field and is hence insensitive to a large proportion of the available modes, but in the case of cosmic shear a tomographic approach still provides useful information on the growth of structure if reasonably accurate redshifts are available~\citep{1999ApJ...522L..21H, 2002PhRvD..65f3001H}. The information content of shear  maps may be boosted further by including small well-measured scales (albeit at the price of increased vulnerability to imperfections in the modelling of baryon feedback in the matter power spectrum~\citealt{2011MNRAS.417.2020S, 2018MNRAS.480.2247C, 2019MNRAS.488.1652H}), and exploiting non-Gaussian information in the signal~\citep[e.g.][]{2002A&A...389L..28B, 2004MNRAS.348..897T, 2010ApJ...712..992B, 2011PhRvD..84d3529Y, 2012MNRAS.423..983P, 2013PhRvD..88l3002P}. As is well known, galaxy weak lensing comes with its own particular observational systematics, for which we refer the reader to~\citet{2018ARA&A..56..393M} for a review.

Given its potential constraining power and the enhanced role that weak lensing is expected to play in shaping our understanding of the Universe, it is timely to ask the following question: which of the various outstanding questions of modern cosmology can lensing, without recourse to other probes, be expected to answer definitively? The power of lensing to constrain models of dark energy, massive neutrinos, and modified gravity has been well documented and demonstrated~\citep[e.g.][]{2008PhRvD..78d3002S, 2009A&A...500..657T, 2010GReGr..42.2177H, 2011PhRvD..83b3012M, 2012JCAP...11..011D, 2013MNRAS.429.2249S, 2015MNRAS.454.2722H, 2017MNRAS.471.1259J, 2019PhRvD..99l3505A} so in this work we focus on the best-fitting cosmological model - flat $\Lambda$CDM, with massive neutrinos. In particular we will pay special attention to the ability of lensing to constrain the Hubble constant, $H_0$. This is an interesting parameter to study with weak lensing for two reasons.

Firstly, there is currently a moderate discrepancy between the value of $H_0$ inferred from Cepheid-calibrated Type-1a supernovae measured by the SH0ES collaboration, which gives $H_0 = 74.03 \pm 1.42 \; \mathrm{km} \, \mathrm{s}^{-1} \, \mathrm{Mpc}^{-1}$, ~\citep{2019ApJ...876...85R}, and that inferred from the primary CMB anisotropies measured by Planck, which gives $67.27 \pm 0.60  \; \mathrm{km} \, \mathrm{s}^{-1} \, \mathrm{Mpc}^{-1}$~\citep{2020A&A...641A...6P} or $67.44 \pm 0.58  \; \mathrm{km} \, \mathrm{s}^{-1} \, \mathrm{Mpc}^{-1}$ with the reanalysis of~\citet{2019arXiv191000483E}. This represents a $4.3$ - $4.4\sigma$ discrepancy known as the `$H_0$ tension', for which there are many proposed solutions~\citep[see, e.g.][for discussion]{2016JCAP...10..019B, 2019NatAs...3..891V, 2021arXiv210301183D}. Having independent measurements of $H_0$ is clearly of great value for determining if the tension is due to undiagnosed systematic errors or genuinely new physics. It is therefore timely to investigate if weak lensing can, or ever will, constrain $H_0$ to useful precision. There is in fact reason to believe that lensing \emph{can} help constrain $H_0$, by combining with baryon acoustic oscillations (BAO). Several works have used large-scale structure probes to constrain $\Omega_m$, which allows $H_0$ to be measured from BAO at a single redshift if the baryon density $\Omega_b h^2$ is constrained a priori, for example from Big Bang Nucleosynthesis (BBN) modelling plus a measurement of the primordial deuterium abundance~\citep[e.g.][]{2013MNRAS.436.1674A, 2018MNRAS.480.3879A, 2019JCAP...10..029S}. Such a measurement of $H_0$ is almost CMB-independent, requiring only the temperature monopole to calibrate the sound horizon. Weak lensing on its own does not constrain $\Omega_m$ well due to a degeneracy with $\sigma_8$ (a degeneracy known as the lensing `banana'), but this degeneracy can be broken by combining galaxy lensing with CMB lensing - the direction of the $\Omega_m$-$\sigma_8$ degeneracy is different for CMB lensing because the redshifts and scales probed are quite different to those of galaxy lensing. Investigating how effective this combination of probes is for constraining $H_0$ is an aim of this work.

Secondly, there have been longstanding difficulties in the interpretation of $H_0$ measurements from lensing due to inherent degeneracies in lensing observables, in particular strong lensing time delays~\citep[e.g.][]{2000AJ....120.1654S, 2002ApJ...578...25K}. This stems from the fact that the dimensionless quantities one can form from lensed images are invariant under a scaling of $H_0$. Dimensionful quantities do change with $H_0$, but this change is degenerate with a redefinition of the (unknown) lens mass density and unlensed source positions~\citep{1985ApJ...289L...1F}. For strong lensing this means that image positions, fluxes, and time delays are invariant under this redefinition and a simultaneous rescaling of $H_0$, known as the `mass sheet degeneracy'~\citep[e.g.][]{2013A&A...559A..37S}. In weak lensing the only effects of changing $H_0$ are to rescale angular diameter distances and the mass density at every point, but given the observables (i.e.~the shear correlation functions) are dimensionless one might wonder if this change can be entirely absorbed by a change in length and mass units. The answer lies in the fact that some external information on the \emph{statistical} lens mass distribution is effectively included through use of a model for the matter power spectrum. Rescaling $H_0$ results in a rescaling of $\Omega_m h^2$, changing the shape of the matter power spectrum as a function of wavenumber and hence changing the angular correlation function of the lensing shear. There do exist weak lensing analysis techniques that try to discard all information from the matter power spectrum, e.g.~shear-ratio tests~\citep{2003PhRvL..91n1302J, 2004ApJ...600...17B, 2005ApJ...635..806Z, 2007MNRAS.374.1377T}, and we note that in these probes the sensitivity to $H_0$ drops out entirely. This discussion suggests there may be simple arguments one can make to extract the $H_0$ dependence of weak lensing observables.

A more general aim of this work is to study \emph{where} information on $\Lambda$CDM parameters comes from in weak lensing analyses. It is well known that cosmic shear constrains the parameter combination $S_8 \propto \sigma_8 \Omega_m^{0.5}$ well, whereas all other parameters are weakly constrained or unconstrained. The precision with which $S_8$ is measured (a few percent in modern lensing surveys) is in stark contrast with $H_0$, which is almost completely unconstrained~\citep{2021A&A...646A.140H}. The sensitivity of shear correlation functions to $S_8$ is often justified~\citep[e.g.][]{2017MNRAS.465.1454H, 2020PASJ...72...16H, 2021A&A...646A.140H} by reference to the work of~\citet{1997ApJ...484..560J}. Although the scales and redshifts used in that work are appropriate for those measured in modern surveys, \citet{1997ApJ...484..560J} used a non-linear prescription for the matter power spectrum that is quite different to the halo models used by modern surveys. We will revisit the origin of the $S_8$ dependence in cosmic shear in the context of the halo model, in addition to paying special attention to $H_0$. To aid this investigation we will also consider information from CMB lensing. CMB lensing constraints currently have tighter error bars than cosmic shear on most combinations of $\Lambda$CDM parameters~\citep{2020A&A...641A...8P, 2019ApJ...884...70W, 2020ApJ...888..119B, 2021MNRAS.500.2250D}, and although the redshifts of the relevant gravitational potentials are quite different there are useful analogies to be drawn between the two probes when it comes to studying the origin of parameter information.

This paper is organised as follows. In Section~\ref{sec:lensing_only} we review current cosmological constraints using weak lensing alone, either from cosmic shear, CMB lensing, or their combination. In Section~\ref{sec:WLcosmodep} we dig into the origin of cosmological constraints from weak lensing, paying particular attention to $S_8$ and $H_0$. In Section~\ref{sec:future} we study whether forthcoming surveys will improve on our understanding of $H_0$ using weak lensing alone, before concluding in Section~\ref{sec:conc}. Finally, in a series of appendices we investigate the sensitivity of our results to various prior and modelling choices and investigate how BAO analyses may be assisted by weak lensing to give a constraint on $H_0$ independent from both the primary CMB fluctuations or the classical distance ladder.

We set $c=1$ throughout unless otherwise stated. We will often use $\omega_m$ and $\Omega_mh^2$ interchangeably, and likewise $\omega_b$ and $\Omega_bh^2$. We define the dimensionless quantity $h$ via $H_0 \equiv 100 \, h \; \mathrm{km} \, \mathrm{s}^{-1} \, \mathrm{Mpc}^{-1}$.

\section{Current cosmological constraints using weak lensing alone}
\label{sec:lensing_only}

Before focusing on $H_0$ we will first briefly review cosmological parameter constraints from weak lensing alone.

We consider two examples of weak lensing data sets: the power spectrum of CMB lensing fluctuations from Planck~\citep{2020A&A...641A...8P}, and the tomographic correlation functions of galaxy ellipticities measured in KV450~\citep{2020A&A...633A..69H}\footnote{Specifically we use cosmic shear measurements from the Kilo-Degree Survey and the VISTA Kilo-Degree Infrared Galaxy Survey~\citep{2015MNRAS.454.3500K, 2019A&A...632A..34W, 2020A&A...633A..69H, 2019A&A...624A..92K}, hereafter referred to as KiDS+VIKING. The KiDS data are processed by THELI~\citep{2013MNRAS.433.2545E} and Astro-WISE~\citep{2013ExA....35....1B, 2017A&A...604A.134D}, and the VIKING data are processed by CASU~\citep{2018MNRAS.474.5459G}. Shears are measured using lensfit~\citep{2013MNRAS.429.2858M, 2019A&A...624A..92K}, and photometric redshifts are obtained from PSF-matched photometry~\citep{2019A&A...632A..34W} and calibrated using external overlapping spectroscopic surveys~\citep{2020A&A...633A..69H}.}. These data probe projected gravitational potentials at high redshift ($0.5 \lesssim z \lesssim 3$) and low redshift ($0.1 \lesssim z \lesssim 0.7$) respectively\footnote{We note that KV450 has been superseded by recent weak lensing results from KiDS-1000~\citep{2021A&A...645A.104A, 2021A&A...645A.105G, 2021A&A...646A.140H} which presents an improved constraint on the lensing amplitude $S_8$ having $~3\%$ precision with lensing alone. We do not expect use of the older data to impact the main results of this paper, since constraints orthogonal to $S_8$ in $\Lambda$CDM models have not improved significantly.}. We will first discuss constraints from the two data sets separately, and then from their combination.

\subsection{Constraints from CMB lensing and galaxy lensing separately}
\label{sec:lensing_solo}

Since the published analyses of these data by the respective collaborations use different parameter priors, we first reanalyse the KV450 likelihood using the same priors as the Planck lensing-only analysis, listed in Table~\ref{tab:priors}. The salient differences from the original KV450 analysis are a broader prior on $h$ and tighter priors on $n_s$ and $\Omega_b h^2$, the latter motivated by standard BBN assuming the three Standard Model neutrino species and the primordial deuterium estimate from~\citet{2018ApJ...855..102C}; we will sometimes explicitly label this choice as ``+BBN''. As shown in Appendix~\ref{app:KV450_vary_priors}, these choices result in negligible differences to the KV450 posterior.

\begin{table*}
  \centering
  \caption{Priors on cosmological parameters used in this work. Square brackets denote uniform priors between indicated limits, otherwise priors are Gaussian with the indicated mean and standard deviation.}
  \label{tab:priors}
  \begin{tabular}{cccccc}
    \hline
    \multicolumn{2}{c}{Planck lensing + BBN} & \multicolumn{2}{c}{KV450 + BBN} & \multicolumn{2}{c}{DES}\\
    \hline
    Parameter & Prior & Parameter & Prior & Parameter & Prior\\
    \hline
    $h$ & $[0.4, 1.0]$ & $h$ & $[0.64, 0.82]$ & $h$ & $[0.55, 0.91]$ \\
    $\Omega_ch^2$ & $[0.001, 0.99]$ & $\Omega_ch^2$ & $[0.01, 0.99]$  & $\Omega_m$ & $[0.1,0.9]$\\
    $\Omega_b h^2$ & $0.0222 \pm 0.0005$  &  $\Omega_b h^2$ & $0.0222 \pm 0.0005$ &  $\Omega_b$ & $[0.03, 0.07]$ \\
    $n_s$ & $0.96 \pm 0.02$ & $n_s$ & $[0.7, 1.3]$ & $n_s$ & $[0.87, 1.07]$\\
    $\log(10^{10}A_s)$ & $[1.61, 3.91]$  &  $\log(10^{10}A_s)$ & $[1.7, 5.0]$ & $10^{9}A_s$ & $[0.5, 5.0]$\\
    \hline
  \end{tabular}
\end{table*}

We sample from the KV450 likelihood using the \textsc{MultiNest}~\citep{2009MNRAS.398.1601F} nested sampling code within the \textsc{MontePython} package~\citep{2018arXiv180407261B}. The linear power spectrum was computed with the Boltzmann code \textsc{class}~\citep{2011ascl.soft06020B}, with non-linear and baryon feedback corrections computed with \textsc{hmcode}~\citep{2015MNRAS.454.1958M}. The posterior parameter constraints and Bayesian credible intervals were computed with \textsc{GetDist}~\citep{2019arXiv191013970L}. Nuisance parameters in KV450 are sampled using the same priors as in the original analysis of~\citet{2020A&A...633A..69H}, and the Planck samples here are the publicly available MCMC chains, with the primordial CMB power spectra marginalised out of the lensing response and $N^{(1)}$ bias as described in~\citet{2020A&A...641A...8P}.

\begin{figure}
  \includegraphics[width=\columnwidth]{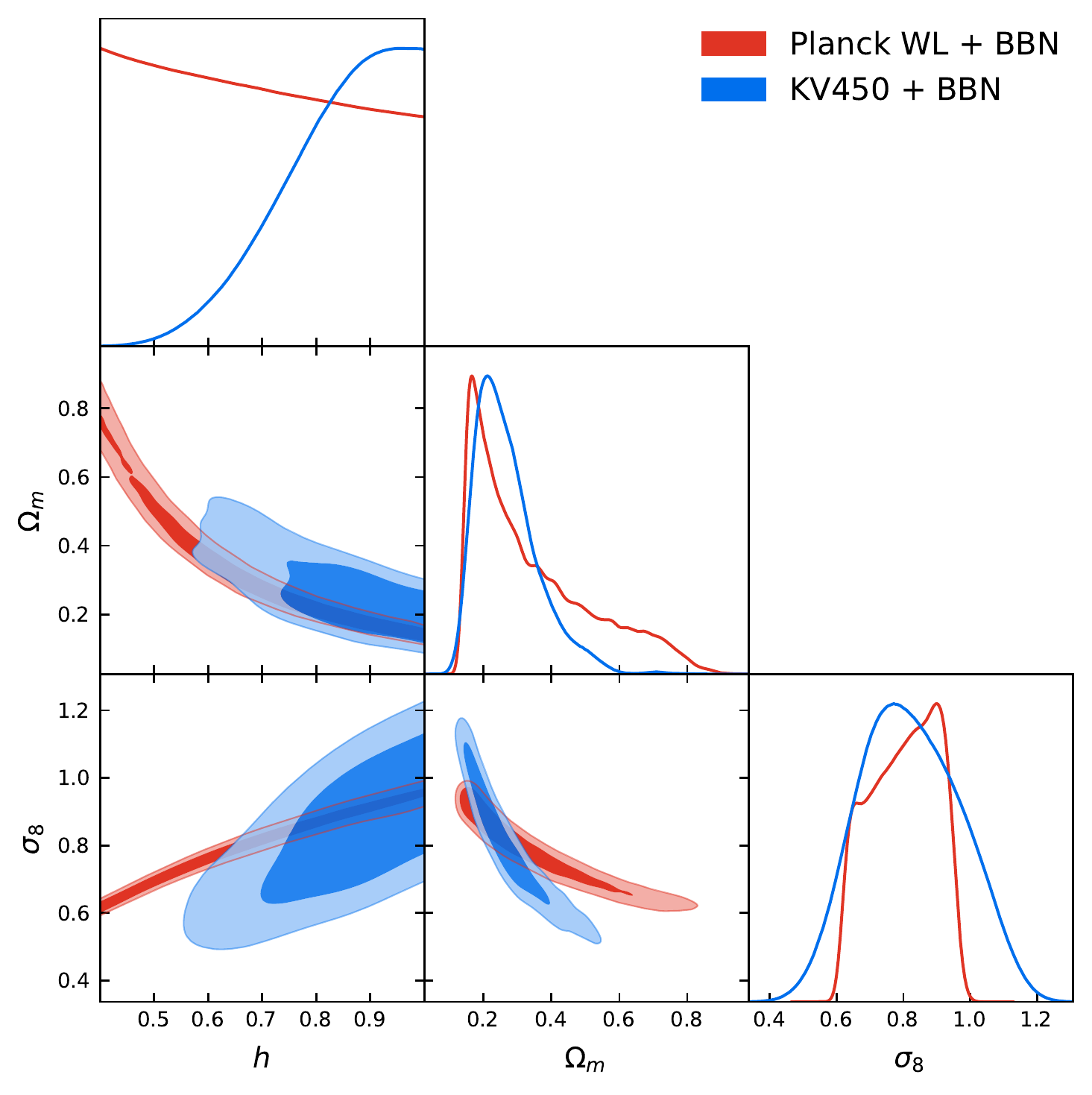}
  \caption{68\% and 95\% posterior credible regions on the parameters $\Omega_m$, $\sigma_8$, and $h$ from the weak lensing correlation functions measured in KV450 (blue) and from the CMB lensing power spectrum measured with Planck (red). The priors used are the `lensing only' priors of~\citet{2020A&A...641A...8P}, i.e.~broad on $h$ and informative on $n_s$ and $\Omega_bh^2$, the latter motivated by BBN.}
  \label{fig:KV450_Pns_BBN_triangle}
\end{figure}

Figure~\ref{fig:KV450_Pns_BBN_triangle} shows constraints from the two lensing-only analyses in the parameter space $\Omega_m$, $\sigma_8$, and $h$. The parameter dependence of CMB lensing has been discussed extensively in~\citet{2014MNRAS.445.2941P, 2016A&A...594A..15P, 2021MNRAS.501.1481H}. The Planck lensing constraints form a `tube' in this parameter space, with two well-constrained parameter combinations roughly corresponding to a measurement of the small-scale amplitude of the lensing power spectrum $C_L^{\phi \phi}$ proportional (at fixed $n_s$ and $\Omega_b h^2$) to $\sigma_8^2 \Omega_m^{-0.05}h^{-1}$ and a measurement of the peak in $[L(L+1)]^2C_L^{\phi \phi}$ given by $L_{{\rm eq}} \propto \Omega_m^{0.6}h$~\citep{2016A&A...594A..15P}. Projected into the traditional `lensing banana' plane of $\sigma_8$ and $\Omega_m$, CMB lensing constrains the combination $\sigma_8 \Omega_m^{0.25}$ (measured with 3\% precision in Planck), which roughly follows from combining the $C_L^{\phi \phi}$ amplitude and peak constraints.

In contrast, the KV450 lensing posterior appears much broader in every parameter plane except $\sigma_8$-$\Omega_m$, where the parameter combination $S_8 \propto \sigma_8 \Omega_m^{0.5}$ is constrained with roughly 5\% precision. The parameter dependence of galaxy lensing has been discussed in~\citet{1997ApJ...484..560J} and will be revisited later, but the difference in well-constrained combinations compared with CMB lensing reflects the different scales and source redshifts probed. CMB lensing probes linear potentials at high redshift, where the growth factor of density fluctuations has not yet been suppressed by $\Lambda$ and the angular diameter distance has non-negligible cosmology dependence - both effects are controlled by $\Omega_m$. These features are in contrast with galaxy lensing, which in addition probes non-linear scales in the matter power spectrum that have a cosmology dependence distinct from that of linear theory. Galaxy lensing also receives a contribution from intrinsic alignments (IAs), which induces further sensitivity to cosmological parameters through its scale and redshift dependence. Despite these differences however, the qualitative degeneracy directions in the two posteriors roughly align in this projected three-parameter space.

That the KV450 posterior is significantly broader than that of Planck in the $\sigma_8$-$h$ and $\Omega_m$-$h$ planes reflects a combination of lower signal-to-noise in the data, posterior broadening from marginalising over nuisance parameters, and lower sensitivity to these parameters in the model. Although $H_0$ in particular is poorly constrained, values of $h\lesssim 0.6$ are clearly disfavoured over $h\gtrsim 0.6$, for example. Figure~\ref{fig:KV450_Pns_BBN_triangle} suggests this reflects a trend in the KV450 posterior to disfavour models with high $\Omega_m$ and low $h$.

\begin{figure*}
  \includegraphics[width=0.8\textwidth]{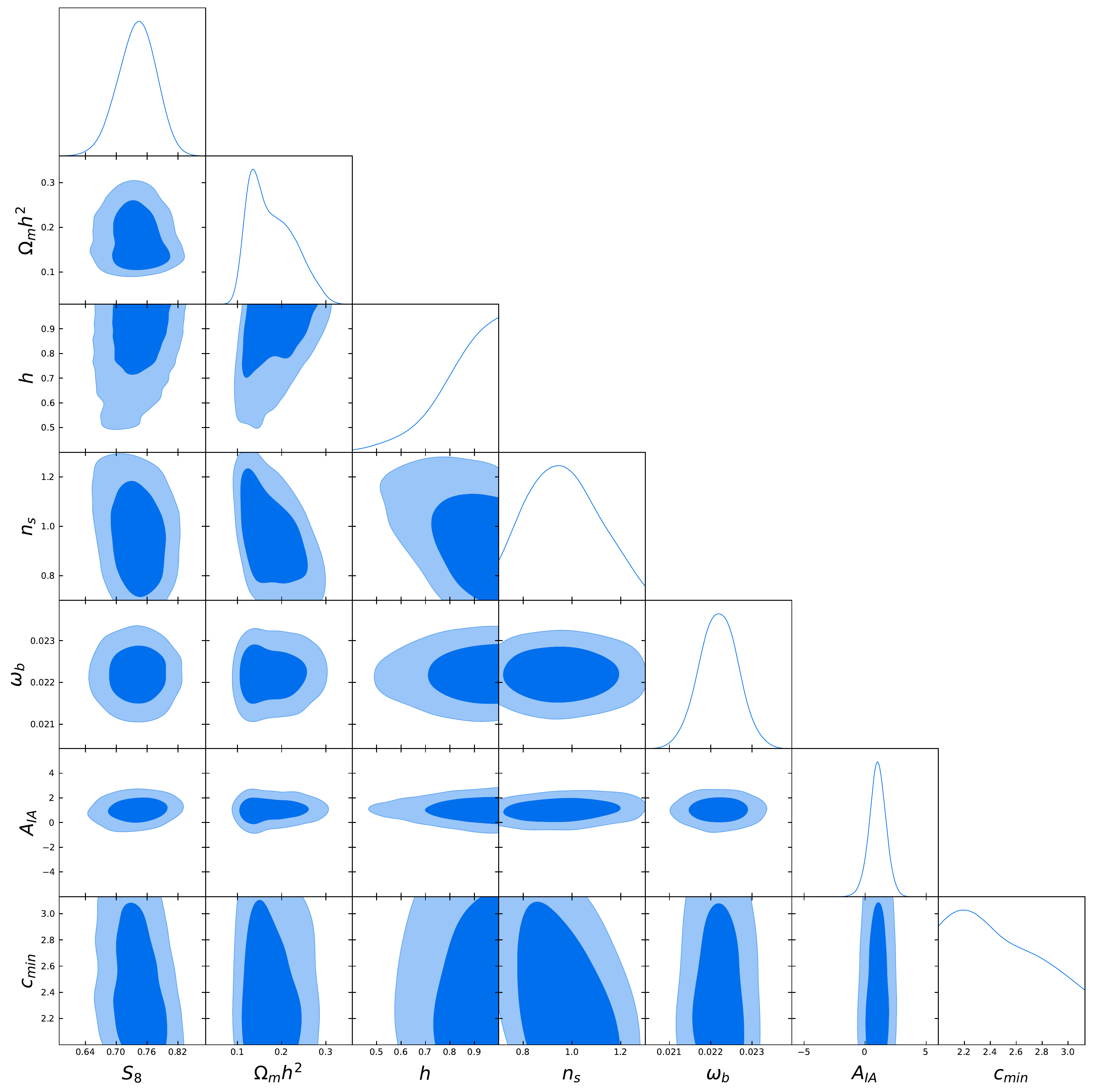}
  \caption{Constraints on cosmological parameters from KV450 with a broad prior on $n_s$ and BBN prior on $\omega_b$. Nuisance parameters have been marginalised over, and we additionally show constraints on the IA amplitude $A_{IA}$ and baryon feedback parameter $c_{{\rm min}}$, whose prior ranges are the same as the plotting range in each case. The value $c_{{\rm min}} = 3.13$ corresponds to no baryon feedback (i.e.~dark matter only), while values smaller than this suppress the matter power spectrum on small scales.}
  \label{fig:KV450_Kns_BBN_cosmoparams}
\end{figure*}

In Figure~\ref{fig:KV450_Kns_BBN_cosmoparams} we show marginalised constraints on all cosmological parameters in the model from KV450 assuming a broad prior on $n_s$ and a BBN prior on $\omega_b$. In contrast with Figure~\ref{fig:KV450_Pns_BBN_triangle} we plot constraints on $[S_8, \Omega_m h^2, h]$, where  $S_8 \equiv \sigma_8 (\Omega_m /0.3)^{0.5}$, instead of $[\sigma_8, \Omega_m, h]$ since the posterior covariance is more diagonal in this basis. All parameters except $S_8$, $A_{IA}$, and $\Omega_m h^2$ are unconstrained by the data (similar statements can be made of DES Y1 cosmic shear~\citealt{2018PhRvD..98d3528T}). The constraint on $\Omega_m h^2$ is weak ($\Omega_m h^2 = 0.177^{+0.035}_{-0.064}$, i.e.~30\% uncertainty) and potentially influenced by the hard priors on $A_s$ and $n_s$, but is orthogonal to $S_8$ and hints at what the more precise forthcoming weak lensing surveys might provide. We will return to this point when we discuss the parameter dependence of galaxy weak lensing.

Thus, in $\Lambda$CDM models with fixed neutrino mass, CMB lensing constrains the parameters $\sigma_8^2h^{-1}$ and $\Omega_m^{0.6}h$ well whereas galaxy lensing constrains $\sigma_8 \Omega_m^{0.5}$ well and $\Omega_m^{0.5} h$ (very) weakly.

\subsection{Constraints from CMB lensing and galaxy lensing combined}
\label{subsec:lensing_joint}

It is clear from Figure~\ref{fig:KV450_Pns_BBN_triangle} that inference of $\Omega_m$, $\sigma_8$, and $h$ from the joint data set will provide improved constraints on both $\sigma_8$ and $\Omega_m$, due to the different degeneracy directions arising from the different source redshifts.

\begin{figure}
  \includegraphics[width=\columnwidth]{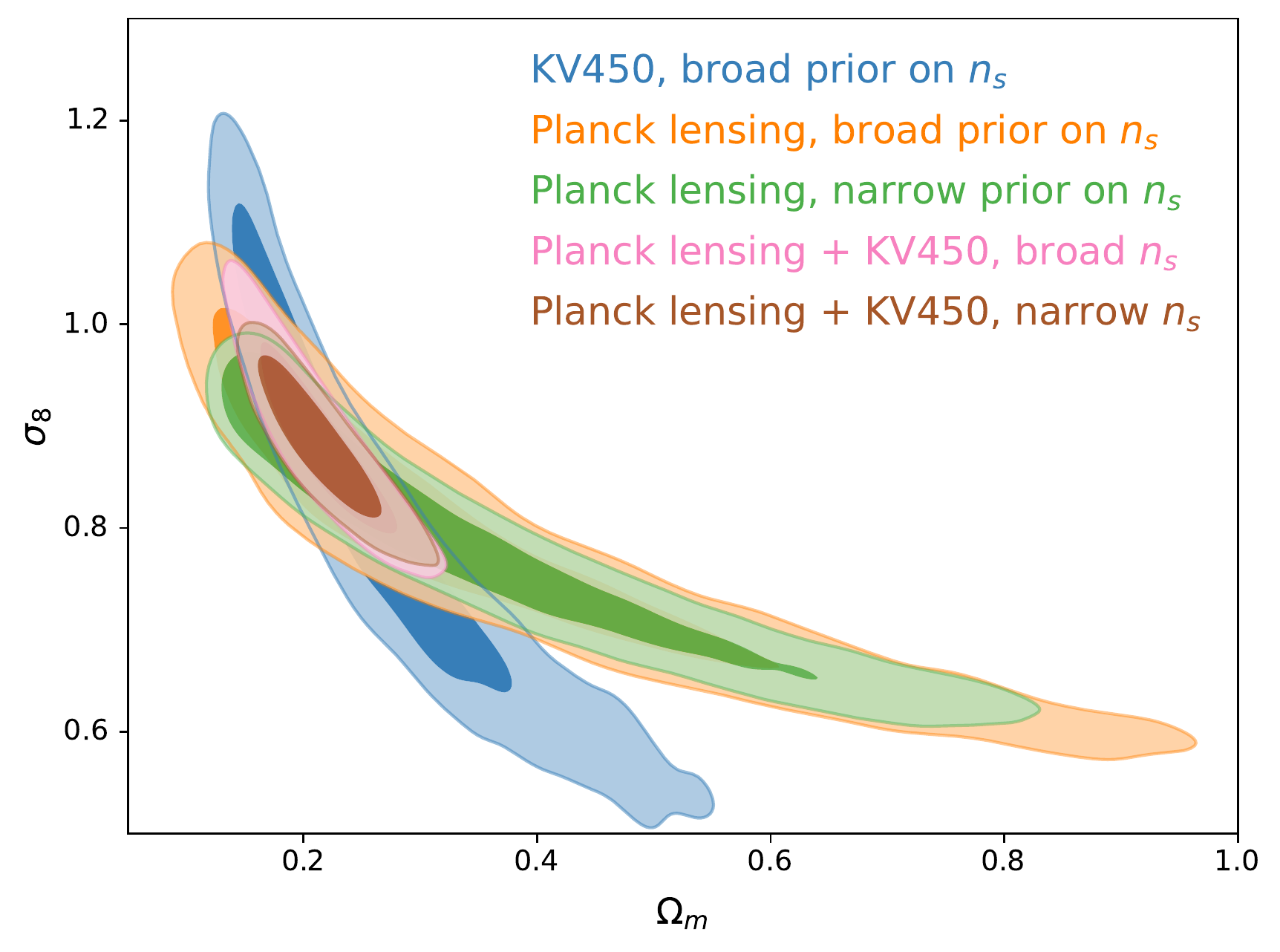}
  \includegraphics[width=\columnwidth]{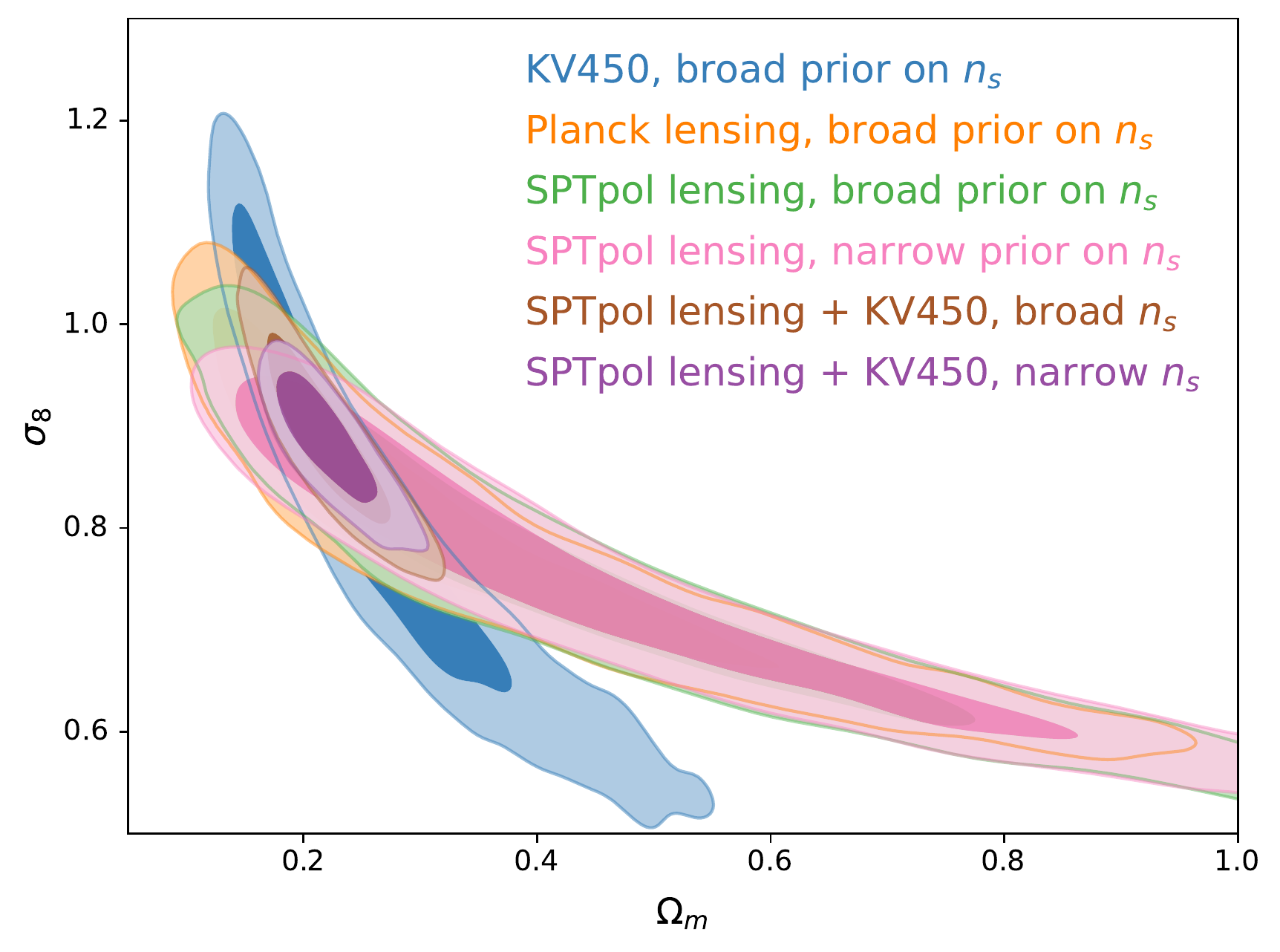}
  \caption{\emph{Top panel}: Marginalised 68\% and 95\% constraints on $\Omega_m$ and $\sigma_8$ from KV450 (blue), Planck lensing (orange and green), and their combination (pink and brown) with either a broad or narrow prior on $n_s$. \emph{Bottom panel}: Same as top, for SPTpol lensing instead of Planck lensing. A BBN prior on $\Omega_b h^2$ has been imposed for each analysis.}
  \label{fig:bananas}
\end{figure}

In the top panel of Figure~\ref{fig:bananas} we show the individual and combined constraints in the $\sigma_8$-$\Omega_m$ plane from Planck and KV450, confirming that this combination provides tighter constraints on both these parameters than in either data set individually. We find $\Omega_m = 0.22 \pm 0.04$ and $\sigma_8 = 0.88 \pm 0.05$, with little sensitivity to the $n_s$ prior, although there is a residual degeneracy between these two parameters. For comparison, combining Planck lensing with DES lensing gives $\Omega_m = 0.266^{+0.041}_{-0.033}$ and $\sigma_8 = 0.837^{+0.042}_{-0.052}$ with `Planck lensing' priors. The preference for higher $\Omega_m$ and lower $\sigma_8$ at the roughly $1\sigma$ level in this combination reflects a slight preference in DES Y1 lensing for higher $S_8$ values, as well as some non-trivial non-Gaussianity in the DES lensing posterior due to intrinsic alignments~\citep{2020A&A...641A...8P}. Planck primary CMB data (TT,TE,EE+lowE) give $\Omega_m = 0.3166 \pm 0.0084$, a discrepancy with Planck lensing + KV450 of about $2.4\sigma$, although we note that $\Omega_m$ and $\sigma_8$ are correlated so the discrepancy is potentially weaker in a higher dimensional parameter space. The preference for lower $\Omega_m$ is not surprising, as it follows from the preference in KV450 for models with lower $S_8$ than Planck primary CMB.

Note that in making Figure~\ref{fig:bananas} we have assumed zero correlation between the CMB lensing and galaxy lensing power spectra. For Gaussian fields the neglected cross-covariance is proportional to the square of the cross-correlation between the two signals. Although this is non-zero (see~\citealt{2014MNRAS.443L.119H} for its modelling and~\citealt{2017MNRAS.471.1619H, 2020arXiv201111613R} for its measurement in KiDS), it is small compared with the diagonal terms of the covariance at each Fourier mode, which justifies treating these the two data vectors as independent.

Cosmological constraints from combining CMB lensing with galaxy lensing have been presented before, for example in~\citet{2020A&A...641A...8P} where Planck lensing was combined with DES galaxy lensing, and more recently in~\citet{2020arXiv200708991E} where redshift-space distortions are additionally included. \citet{2020arXiv201016416T} present constraints from Planck CMB lensing combined with KiDS-1000 galaxy lensing, finding $\Omega_m = 0.269^{+0.026}_{-0.029}$ and $\sigma_8 = 0.81^{+0.047}_{-0.029}$, with a residual degeneracy between these two parameters. These constraints are both roughly $1\sigma$ away from our measurement along their degeneracy direction. This is partly due to the tighter priors on $h$ adopted in KiDS-1000; adopting these priors in our analysis gives better agreement, with $\Omega_m = 0.23^{+0.02}_{-0.03}$ and $\sigma_8 = 0.86\pm 0.04$. A plot similar to Figure~\ref{fig:bananas} also appears in~\citet{2020ApJ...888..119B}, although without constraints from the combination of the data sets. Comparatively little attention has been paid to the sensitivity of such constraints to the assumed priors and to the specific data sets entering the combination. In the top panel of Figure~\ref{fig:bananas} we show how the constraints change when the baseline Planck lensing-only priors are relaxed. Keeping the informative BBN prior on $\Omega_b h^2$, the only remaining informative prior is that on the scalar spectral index $n_s$, which we relax to its less informative KV450 prior, see Table~\ref{tab:priors}. Relaxing the prior on $n_s$ broadens the CMB lensing parameter contours significantly, since $n_s$ can now compensate for values of $\sigma_8$, $\Omega_m$, and $h$ that previously gave rise to power spectrum amplitudes and peaks not favoured by the data. In contrast the galaxy lensing posterior does not change significantly when relaxing the prior on $n_s$, as shown in Figure~\ref{fig:KV450_3params}, suggesting that these new degeneracies between cosmological parameters are subdominant to noise and the broadening of contours arising from marginalising over nuisance parameters. The constraint on $S_8$ is dominated by the KV450 data\footnote{The relative impact of CMB lensing on the $S_8$ constraint is greater in KiDS-1000~\citep{2020arXiv201016416T}, likely due to their tighter prior on $h$ favouring a region of parameter space where the galaxy and CMB lensing contours are more orthogonal.} and is hence more stable to changing the $n_s$ prior, giving $S_8 = 0.742 \pm 0.030$ (narrow prior on $n_s$) and $S_8 = 0.744 \pm 0.028$ (broad prior on $n_s$), i.e. a modest improvement from a 5\% measurement in KV450 alone to 4\% when combined with Planck lensing.

In the bottom panel of Figure~\ref{fig:bananas} we show the constraints in the $\Omega_m$-$\sigma_8$ plane when swapping the Planck CMB lensing measurement with that of SPTpol~\citep{2019ApJ...884...70W, 2020ApJ...888..119B}. We use the SPTpol likelihood of~\citet{2020JCAP...08..013C}, verifying that our results agree with those of~\citet{2020ApJ...888..119B} when adopting the same priors (which are identical to the Planck lensing only priors listed in Table~\ref{tab:priors}). The SPTpol lensing constraints appear highly consistent with those of Planck when projected into this parameter space, but are broader due to the different scales probed; as discussed in~\citet{2020ApJ...888..119B}, SPTpol is not as sensitive as Planck to the large-scale break in the lensing power spectrum, and hence struggles to distinguish $A_s$ from $\Omega_m h^2$, both of which change the small-scale amplitude of lensing. This leads to weaker constraints in the $\Omega_m$-$h$ plane (which is where the peak information is most manifest), and a longer tail to high $\Omega_m$ in the $\sigma_8$-$\Omega_m$ plane. The lensing power spectrum amplitude is still well measured in SPTpol across a wide range of scales, so the parameter $\sigma_8 \Omega_m^{0.25}$ is still tightly constrained with 4\% precision. The combination with KV450 (again neglecting covariance) gives constraints in the $S_8$ direction that are again dominated by KV450, with $S_8 = 0.756\pm 0.028$ (narrow $n_s$ prior) and $S_8 = 0.756\pm 0.027$ (broad $n_s$ prior), i.e. $\sim 4\%$ measurements. The combination constrains $\Omega_m = 0.22 \pm 0.03$ and $\sigma_8 = 0.88 \pm 0.04$ with a narrow prior on $n_s$, broadening to  $\Omega_m = 0.22 \pm 0.04$ and $\sigma_8 = 0.89 \pm 0.06$ with a broad prior on $n_s$. As with Planck+KV450, there is a residual degeneracy between these two parameters even in the combined data set. Note that this low value of $\Omega_m$ implies a high value of $H_0$ when fixing the angular scale of the CMB acoustic peaks to the measurement from Planck, since this essentially fixes $\Omega_m h^3$~\citep{2020A&A...641A...6P}. The naive combination implies roughly $h = 0.76 \pm 0.04$, consistent with the SH0ES value and 2$\sigma$ higher than the published Planck value. This highlights the interplay between the $S8$ and $H_0$ `tensions', a point also discussed in~\citet{PhysRevD.102.043507, 2020arXiv200900006N, 2020arXiv201004158J, 2020PhRvD.102j3502I, 2020JCAP...05..005D, 2020arXiv200612420D, 2021MNRAS.501.1481H}.

Joint parameter inference from the combination of CMB and galaxy lensing can thus offer improved constraints on $\Omega_m$ and $\sigma_8$ through the breaking of their degeneracy. In Appendix~\ref{sec:lensing_bao} we investigate whether this improved constraint on $\Omega_m$ can be used to calibrate the sound horizon (in combination with a BBN measurement of $\Omega_bh^2$) in order to measure $H_0$ in combination with BAO. Despite the tighter $\Omega_m$ constraint, constraints in the $\Omega_m$-$H_0$ plane are not improved substantially, with $\Omega_m^{0.6}h$ still a degeneracy direction for this data combination. The reason for this is the fairly weak dependence of the equality angular scale $L_{\mathrm{eq}}$ on $\Omega_m$ combined with an improvement in the $\Omega_m$ constraint that is only modest. Low redshift BAO+BBN give a constraint in the $\Omega_m$-$H_0$ plane that is roughly orthogonal to that from CMB lensing, so the additional coarse information on $\Omega_m$ from galaxy lensing does not improve the constraint on $H_0$ significantly. The tightest constraint comes from combining low and high redshift BAO with CMB and galaxy lensing, which gives $H_0 = 67.4 \pm 0.9 \; \mathrm{km} \, \mathrm{s}^{-1} \, \mathrm{Mpc}^{-1}$ for Planck lensing priors + BBN, and  $H_0 = 67.6 \pm 1.1 \; \mathrm{km} \, \mathrm{s}^{-1} \, \mathrm{Mpc}^{-1}$ for KV450 lensing priors + BBN. These are $4.0\sigma$ and $3.6\sigma$ lower than the SH0ES value respectively.

Our BBN prior on the baryon density depends on the assumption that the effective number of relativistic degrees of freedom, $N_{{\rm eff}}$, takes its standard value $N_{{\rm eff}} = 3.046$~\citep[e.g.][]{2018ApJ...855..102C}. To further decouple our analysis from early-Universe physics, we experimented with discarding the BBN prior entirely. Keeping the informative $n_s$ prior, our full lensing + BAO combination yields $H_0 = 70.0^{+8.4}_{-4.6} \; \mathrm{km} \, \mathrm{s}^{-1} \, \mathrm{Mpc}^{-1}$, consistent with both Planck and SH0ES. On the lensing side this constraint is dominated by the Planck CMB lensing measurement of the angular size of the matter-radiation equality scale, with the uncalibrated BAO providing a measurement of $\Omega_m$ that breaks the degeneracy with $H_0$~\citep[see][for a discussion of how uncalibrated standard rulers can be used to constrain $\Omega_m$]{2021arXiv210205701L}. Note that we are still assuming $N_{{\rm eff}} = 3.046$, such that the equality scale is controlled purely by $\Omega_m h^2$ at fixed CMB temperature. Relaxing this would likely destroy almost all the $H_0$-constraining power of lensing.

\section{Dependence of cosmic shear on $H_0$}
\label{sec:WLcosmodep}

We have seen that current galaxy weak lensing data are not powerful enough on their own to improve $H_0$ constraints significantly. Given that CMB lensing gives comparatively tight constraints in the $\Omega_m$-$h$ plane with only a single source redshift, it is natural to ask what is causing cosmic shear to be so poor at providing useful $H_0$ information. While the effects of noise and nuisance parameters certainly contribute, it is interesting to note that constraints on the amplitude of the lensing power spectrum (through $S_8$) are comparable to that of CMB lensing, suggesting that these effects are either relatively more important in the $H_0$ direction or are subdominant to a potential loss of sensitivity in the model to $H_0$ over the scales and redshifts probed. Given that forthcoming lensing surveys will have significantly lower statistical noise, this motivates a more detailed investigation into the cosmology dependence of cosmic shear two-point functions.

\citet{1997ApJ...484..560J} (hereafter~\citetalias{1997ApJ...484..560J}) studied the dependence of the shear correlation function on cosmological parameters in the linear and non-linear regime. \citetalias{1997ApJ...484..560J} found that the amplitude of the correlations in flat $\Lambda$CDM models scales as $\sigma_8 \Omega_m^\alpha$, with $\alpha \lesssim 0.5$ on angular scales $\theta \lesssim 2'$ and $\alpha \approx 0.7$ on scales $\theta > 10'$, with additional (albeit sub-linear) dependencies on the source redshift. The amplitude scaling was found to be only weakly sensitive to the shape of the matter power spectrum (i.e. $n_s$ and $\Gamma \equiv \Omega_m h$). In deriving these scalings, \citetalias{1997ApJ...484..560J} applied a correction to the matter power spectrum for non-linear growth from~\citet{1996MNRAS.280L..19P}, showing that non-linear evolution has an important role in dictating the scaling of the lensing amplitude with cosmological parameters. The scales and redshifts probed by modern lensing surveys is such that the $\sigma_8 \Omega_m^{0.5}$ dependence predicted by \citetalias{1997ApJ...484..560J} closely approximates the actual parameter combination best constrained by the data, so much so that this combination has been given its own name, $S_8$.

There are two points to make about this successful prediction for the lensing amplitude scaling. Firstly, as pointed out by~\citetalias{1997ApJ...484..560J} the scaling with $\Omega_m$ and $\sigma_8$ changes with angular scale, with a stronger dependence on $\Omega_m$ seen on large linear scales. Forthcoming weak lensing surveys will have enough sky area to measure precise shear correlations on degree scales and larger, suggesting that the best-constrained parameter combination from lensing may soon differ from $S_8$. Secondly, the $S_8$ scaling derived in~\citetalias{1997ApJ...484..560J} was made using an approximate non-linear model that has been largely superseded in weak lensing by the halo model and its variants~\citep{2000MNRAS.318..203S, 2000MNRAS.318.1144P, 2002PhR...372....1C}. Given this discrepancy, it is worth revisiting the arguments of~\citetalias{1997ApJ...484..560J} in the context of the halo model. In particular, the parameter $H_0$ appears nowhere in the scalings presented by~\citetalias{1997ApJ...484..560J}. We will therefore pay special attention to how (if at all) cosmic shear correlations depend on $H_0$, and why $S_8$ continues to be the only well-measured $\Lambda$CDM parameter combination in modern lensing analyses.

The cosmological constraints from galaxy lensing we have presented are derived from measurements of the shear correlation functions $\xi_{+/-}^{ij}(\theta)$ between pairs of galaxies in tomographic redshift bins $i$ and $j$ separated by an angle $\theta$. In the flat-sky approximation and neglecting B-modes (e.g.,~\citealt{2015RPPh...78h6901K}) these are given by
\begin{equation}
  \xi_{+/-}^{ij}(\theta) = \int \frac{\ell \mathrm{d}\ell}{2\pi} C_\ell^{ij} J_{0/4}(\ell \theta),
  \label{eq:xipm}
\end{equation}
where $J_n(x)$ is a Bessel function and $C_\ell^{ij}$ is the angular power spectrum of the ellipticity E-mode between tomographic bin pair [$i$, $j$]. Equation~\eqref{eq:xipm} is then subsequently averaged (with an appropriate weighting) over $\theta$ within angular bins. The cosmology dependence of the correlation functions thus follows from that of $C_\ell$, with the scale dependence mapped to that of the data by the bin-averaged Bessel functions.

\begin{figure}
  \includegraphics[width=\columnwidth]{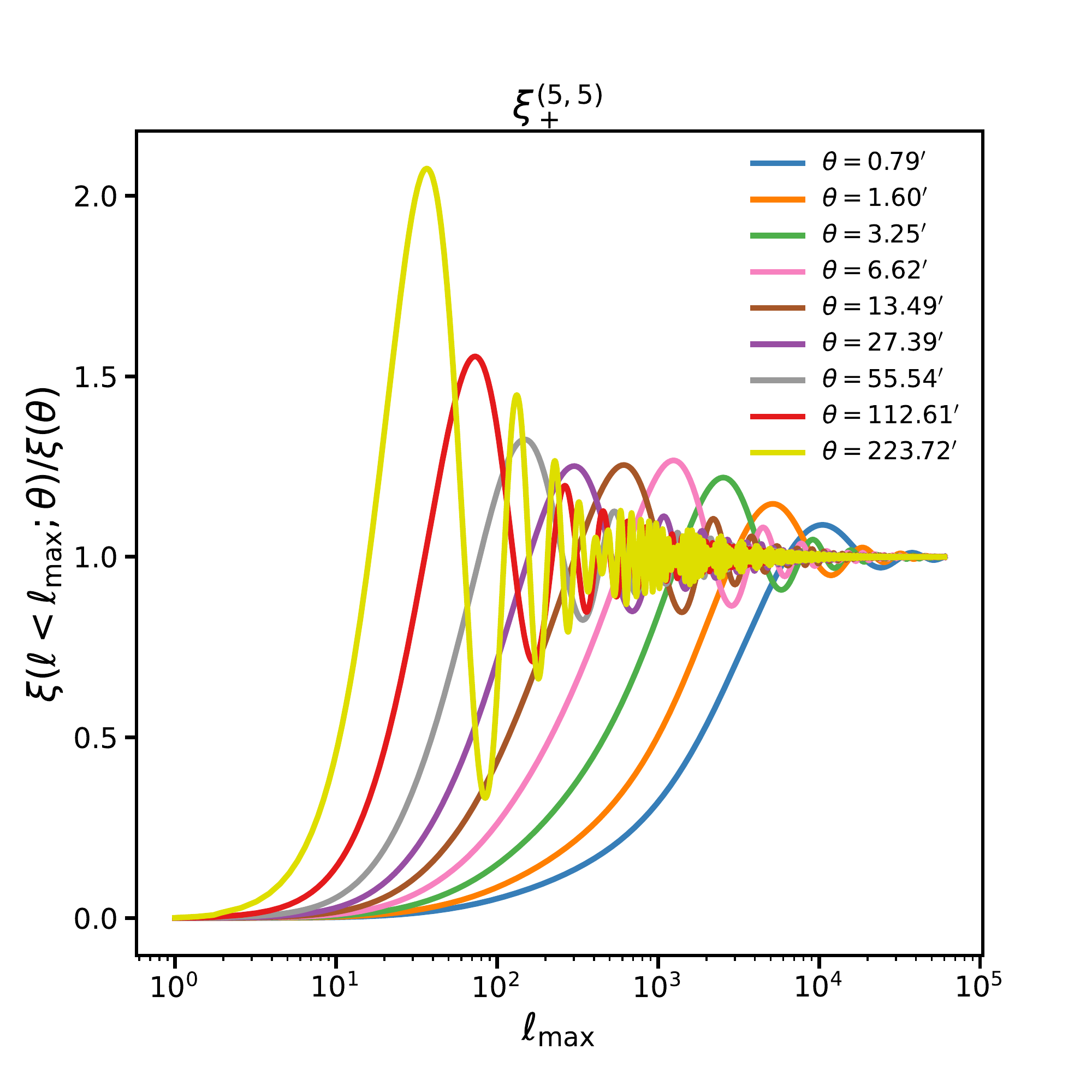}
  \includegraphics[width=\columnwidth]{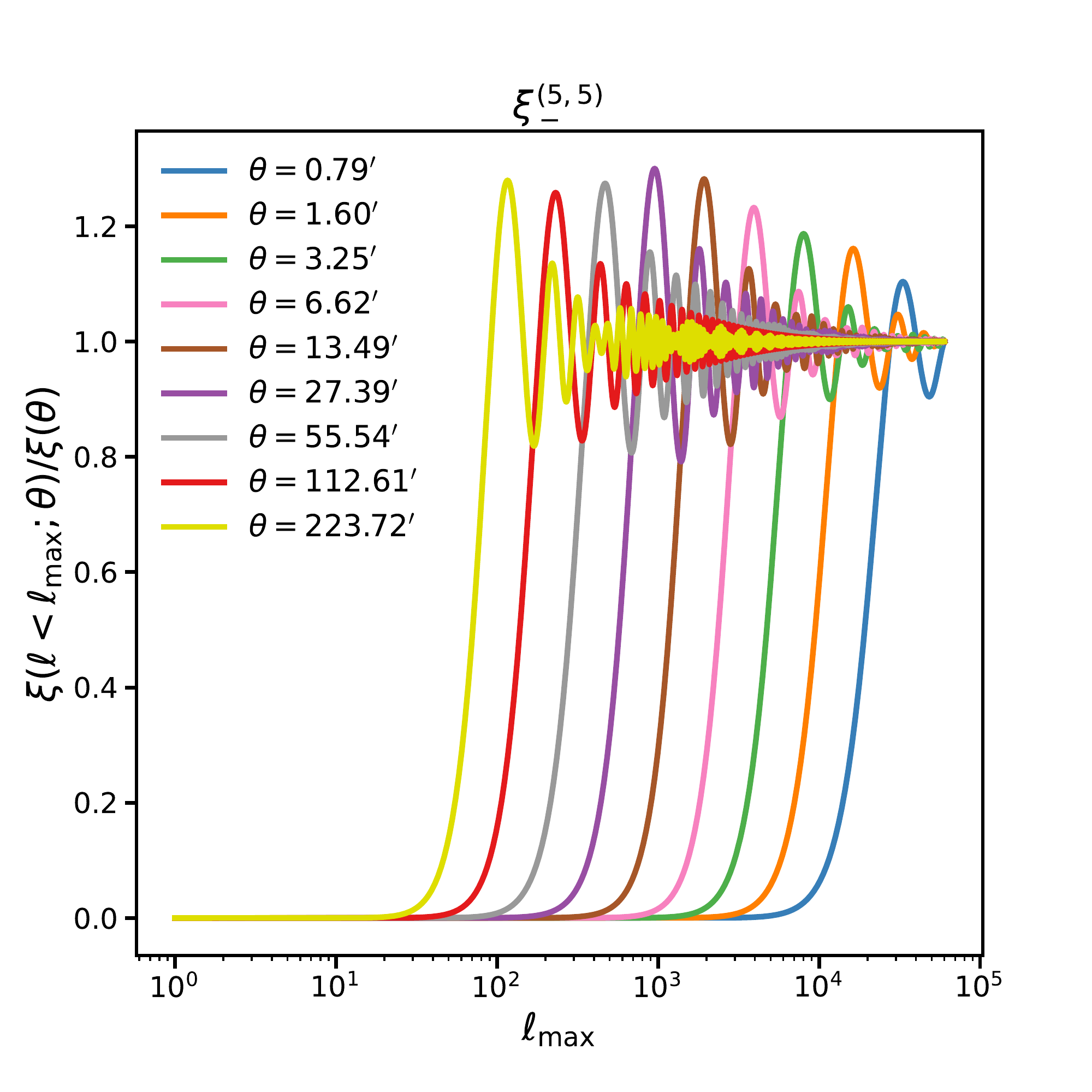}
  \caption{Cumulative contribution per angular wavenumber to the KV450 lensing correlation functions for the angular bins labelled in the top panel (corresponding to those actually measured). The upper panel shows the $\xi_+$ autocorrelation for bin 5 (central redshift $z=1$, top row), the lower panel shows the $\xi_-$ autocorrelation. The curves have been normalised by their values when all multipoles are included. Other bin combinations follow similar trends. The $\xi_-$ Fourier kernel is much more localised in $\ell$ than that of $\xi_+$, which explains qualitatively the difference between these plots (see, e.g., Figure 1 of~\citealt{2021A&A...645A.104A} for an un-integrated version of this plot).}
  \label{fig:KV450_xi_kernels}
\end{figure}

In Figure~\ref{fig:KV450_xi_kernels} we plot the contribution per-$\ell$ to the shear correlation functions in the KV450 angular bins, for a few different redshift bin combinations. The sensitivity to each $\ell$ is additionally set by the covariance matrix of $\xi_{+/-}^{ij}(\theta)$, but this plot shows that the KV450 correlation functions probe a broad range of scales $10^2 \lesssim \ell \lesssim 10^4$.

The ellipticity angular power spectrum has contributions from shear and intrinsic alignments, and can be written as
\begin{equation}
  C_\ell^{ij} = C_\ell^{\gamma \gamma, ij} +  C_\ell^{\gamma I, ij} +  C_\ell^{I \gamma, ij} +  C_\ell^{II, ij},
  \label{eq:Clee}
\end{equation}
where both cross-terms are included to allow for overlapping source redshift bins. In the Limber and Born approximations the shear power spectrum is given by
\begin{equation}
  C_\ell^{\gamma \gamma, ij} = \frac{9}{4}\Omega_m^2 H_0^4 \int_0^{r_{{\rm max}}} \mathrm{d}r \, \frac{q_i(r)q_j(r)}{r^2} P\left(\frac{\ell + 1/2}{r}; z(r)\right),
  \label{eq:limber}
\end{equation}
where $P(k; z)$ is the matter power spectrum at redshift $z$, $r_{\mathrm{max}}$ is a maximum source conformal distance, and in flat models
\begin{equation}
  q_i(r) = \frac{r}{a(r)} \int_r^{r_{{\rm max}}} \mathrm{d}r' \, p_r^i(r') \frac{r' - r}{r'},
\end{equation}
where $p_r^i(r)$ is the source density of bin $i$ in $r$-space and integrates to unity, and $a(r)$ is the scale factor at conformal distance $r$ on the background light-cone, normalized as $a(0) = 1$.

Following the fiducial KV450 analysis we adopt the `non-linear linear alignment' model for IAs~\citep{2004PhRvD..70f3526H, 2007NJPh....9..444B} such that
\begin{equation}
  C_\ell^{II, ij} = \int_0^{r_{{\rm max}}} \mathrm{d}r \, \frac{p_r^i(r)p_r^j(r)}{r^2} A_{IA}^2(r) \, P\left(\frac{\ell + 1/2}{r}; z(r)\right),
\end{equation}
where
\begin{equation}
  A_{IA}(r) = -A_{IA}  \frac{5 \times 10^{-4}}{h^2} \frac{\rho_c^0}{M_\odot \,  \mathrm{Mpc}^{-3}} \frac{\Omega_m}{D(z)/D(0)} \left( \frac{1+z}{1+z_0} \right)^\eta,
  \label{eq:AIAr}
\end{equation}
where $\rho_c^0$ is the critical density and $D(z)$ is the matter density growth factor (assuming scale-independent growth). An additional redshift dependence arises when $\eta \neq 0$, but following KV450 we set $\eta=0$ in our analysis. The factor $A_{IA}$ is a dimensionless scaling amplitude with fiducial value of unity. The GI terms in Equation~\eqref{eq:Clee} follow from the cross terms given in, e.g.,~\citet{2017MNRAS.465.1454H}.

\subsection{Dependence of the shear angular power spectrum on $H_0$}
\label{subsec:Clcosmodep}

How does the lensing power spectrum depend on $H_0$? To answer this, we will first focus on $C_\ell^{\gamma \gamma}$, given in Equation~\eqref{eq:limber}. Consider fixing $\Omega_m$ and scaling $H_0$ by a constant factor $\alpha$ such that
\begin{equation}
  h \rightarrow h(1+\alpha).
  \label{eq:alpha}
\end{equation}
Changing integration variables to the $\alpha$-independent quantity $r_h \equiv r h$ and defining $L \equiv \ell+1$ we have
\begin{equation}
  C_\ell^{\gamma \gamma, ij} \propto L^{-3} \Omega_m^2 \int_0^{r_{h,{\rm max}}} \mathrm{d}r_h \frac{q_i(r_h)q_j(r_h)}{r_h^2} r_h^3 \Delta_h^2\left( \frac{L}{r_h}; z(r_h) \right),
  \label{eq:Clgg_scaled}
\end{equation}
up to numerical constants, where the function $\Delta^2_h(x)$ is the usual dimensionless matter power spectrum as a function of $k/h$, i.e. $\Delta^2_h(k/h) \equiv \Delta^2(k)$ with $\Delta^2(k) = k^3P(k)/2\pi^2$.

The functions $q_i(r_h)$ are invariant under the $\alpha$-scaling (and hence independent of $h$), and in $\Lambda$CDM are purely functions of $\Omega_m$. From Equation~\eqref{eq:AIAr} we see that the intrinsic alignment amplitude is also invariant under the $\alpha$-scaling, and an expression analogous to Equation~\eqref{eq:Clgg_scaled} can be written down for the $II$ and $GI$ terms. This shows that the dependence of lensing observables on $h$ is entirely captured by its influence on the dimensionless matter power spectrum expressed in ${\rm Mpc}/h$ length units, or alternatively on the dimensionful power spectrum $P(k)$ in units of $h^{-3} \, {\rm Mpc}^3$ and with $k$ in $h \, {\rm Mpc}^{-1}$ units (i.e. the form in which it is usually presented). This should come as no surprise of course since the equations of lensing can equally well be derived from a conformally-rescaled spacetime metric, i.e. no preferred length scale is introduced\footnote{The same effect gives rise to the mass sheet degeneracy familiar from strong lensing. Indeed, the scaling in Equation~\eqref{eq:alpha} corresponds to scaling the mean density of the Universe by $(1+\alpha)^2$.}.

Dependence on $H_0$ is thus imparted to the lensing power spectrum through length scales present in the matter power spectrum. The most relevant scales will be features (peaks, breaks) in the matter power spectrum that survive the lensing projection, so we can expect scales such as the matter-radiation horizon scale, the non-linear transition scale, and scales associated with smaller scale halo structure to be the most important.

\subsection{Dependence of the matter power spectrum on $H_0$ in the halo model}
\label{subsec:PKcosmodep}

The dependence of lensing on $H_0$ follows from that of the matter power spectrum, so we now study the effects of $H_0$ in $\Delta^2(k)$. Since non-linear scales determine much of the cosmological sensitivity of present weak lensing surveys, we follow the fiducial KV450 analysis and use \textsc{hmcode}~\citep{2015MNRAS.454.1958M} as our prescription for non-linear matter clustering. This augments the original model of~\citet{2000MNRAS.318..203S, 2000MNRAS.318.1144P} with updated prescriptions for the fundamental building blocks of the halo model (the halo mass function, the density profile, and the halo bias) calibrated against N-body simulations. The $\Lambda$CDM cosmology dependence of the resulting matter power spectrum agrees with the accurate emulator-based results of~\citet{2014ApJ...780..111H} to a few percent down to $k \approx 10 \, h \, \mathrm{Mpc}^{-1}$ at $z=0$ (see Figure A1 of~\citealt{2015MNRAS.454.1958M}).

In the halo model, the matter power spectrum is written as\footnote{Note that in \textsc{hmcode} this expression is modified slightly to improve the modelling of the transition region.}
\begin{equation}
  \Delta^2(k) = \Delta^2_{2H}(k) + \Delta^2_{1H}(k),
\end{equation}
where $\Delta^2_{2H}$ and $\Delta^2_{1H}$ are the 2-halo and 1-halo terms respectively. We consider the $h$-dependence of these terms separately.

\subsubsection{2-halo term}
\label{subsubsec:2halo}

At current weak lensing precision it is a good approximation to set the 2-halo term equal to the linear matter power spectrum, $\Delta_L^2$. Following the notation of~\citet{2016A&A...594A..15P} and taking $n_s=1$, we write the asymptotic limits of the linear power spectrum at $z=0$ as
\begin{equation}
  \Delta_L^2(k) \propto
  \begin{cases}
    A_s  g^2(a=1) \frac{k^4}{\omega_m^2} & k \ll k_{\mathrm{eq}} \\[10pt]
    A_s  g^2(a=1) \frac{k_{\mathrm{eq}}^4}{\omega_m^2} \ln^2(k/k_{\mathrm{eq}})\; \alpha_\Gamma^2(\omega_m, \omega_b) & k \gg k_{\mathrm{eq}}
    \label{eq:linear_model}
  \end{cases}
\end{equation}
where $k_{\mathrm{eq}} \propto \omega_m$ is the matter-radiation equality scale and $g(a)$ is the growth factor of the Newtonian potential normalised to unity at high redshift. In $\Lambda$CDM models this depends on parameters as $g(a=1) \propto \Omega_m^{0.23}$ in the vicinity of $\Omega_m = 0.3$, as found in~\citet{2016A&A...594A..15P}. The function $\alpha_\Gamma$ accounts for the effects of baryons on the matter transfer function; prior to decoupling from the photons, baryons cannot cluster on scales below the sound horizon due to the pressure support of the photon-baryon plasma. This suppresses the gravitational potential relative to what it would be if all the matter were in CDM~\citep{1998ApJ...496..605E}. Note that we have assumed small deviations around the best-fit $\Lambda$CDM cosmology, such that the limit $k \gg k_{\mathrm{eq}}$ also corresponds to $kr_{{\rm drag}} \gg 1$, where $r_{{\rm drag}}$ is the sound horizon when the baryons decouple. Note also that we have neglected the BAO wiggles, as well as Silk damping and other subdominant effects. From Equation 31 of~\citet{1998ApJ...496..605E} we find $\alpha_{\Gamma} \propto \omega_m^{0.36} \omega_b^{-0.20}$ in the vicinity of our fiducial model. Since we fix $\omega_b$ throughout this section, this results in an extra factor of $\omega_m^{0.72}$ in the power spectrum in the limit that $kr_{{\rm drag}} \gg 1$.

We will express the amplitude of the linear power spectrum in terms of $\sigma_8$. In the vicinity of our model with $n_s \approx 1$ this depends on parameters as~\citep{2016A&A...594A..15P}
\begin{align}
  \sigma_8^2 & \propto \, A_s \Omega_m^{0.46} \omega_m^{1.05} h^{1.4} \nonumber \\
  & \propto \, A_s \Omega_m^{1.51} h^{3.5},
\end{align}
where the $h$-dependence follows from defining the variance in a sphere $\sigma^2(R)$ in terms of a fixed $Rh$, with an additional dependence at fixed $\Omega_m$ from the shape of the power spectrum through $\omega_m$. More generally, in the vicinity of $R = 8 \, h^{-1} \, \mathrm{Mpc}$, we have $\sigma(R) \approx \sigma_8 [R/(8 \, h^{-1} \, \mathrm{Mpc})]^{-0.7}$ with the steepness of this relation increasing with $R$. This will be important when we discuss the 1-halo term.

At $z=0$ we can thus approximate the linear power spectrum (taking $n_s = 1$) as
\begin{equation}
  \Delta_L^2(k) \propto
  \begin{cases}
    \sigma_8^2 \Omega_m^{-3.05} h^{-3.5} (k/h)^4 & k \ll k_{\mathrm{eq}}, z=0 \\[10pt]
    \sigma_8^2 \Omega_m^{1.67} h^{1.94} \ln^2\left(\frac{k/h}{\Omega_m h} \right) & k \gg k_{\mathrm{eq}}, z=0.
  \end{cases}
  \label{eq:D2lin}
\end{equation}
At high redshift the potential growth factor tends to unity, so the parameter dependence of $\Delta_L^2$ can be found by multiplying Equation~\eqref{eq:D2lin} by $\Omega_m^{-0.46}$.

We thus see that, at fixed $\sigma_8$, the amplitude of the linear power spectrum depends strongly on $\Omega_m$ and $h$ on scales larger than the equality scale, with the dependence on $h$ dramatically weakening on smaller scales. The break scale in the linear $\Delta_h^2$ occurs at $k_{\mathrm{eq}}/h \propto \Omega_m h$, which immediately suggests that if a lensing experiment can measure this break then an accurate constraint on $\Omega_m^{\alpha} h$ will be possible, where $\alpha < 1$ due to the $\Omega_m$-dependence of the comoving angular diameter distance. This is exactly what causes CMB lensing to be sensitive to the parameter combination of $\Omega_m^{0.6} h$. Current galaxy lensing surveys however are not wide enough to measure the equality scale, which immediately hints at why $h$ is so poorly constrained in current cosmic shear surveys.

From Equation~\eqref{eq:Clgg_scaled} we see that the lensing power spectrum receives an extra factor of $\Omega_m^{2+\epsilon}$, where $\epsilon < 0$ arises from the cosmology dependence of the comoving angular diameter distance. The latter is weak given the low redshifts of the lenses probed by current galaxy surveys, so we ignore it in these estimates. On large angular scales therefore we can expect $C_\ell \sim \sigma_8^2 \Omega_m^{-1} h^{-3.5}$, transitioning to $C_\ell \sim \sigma_8^2 \Omega_m^{3.7} h^{1.9}$ on small (but still linear) scales with the transition at $\ell_{\mathrm{eq}} \approx \Omega_m h$.

\subsubsection{1-halo term}
\label{subsubsec:1halo}

The 1-halo term at any redshift is written as
\begin{equation}
  \Delta^2_{1H}(k) = \frac{k^3}{2\pi^2} \int_0^\infty \mathrm{d}M \, n(M) W^2(k,M),
  \label{eq:1hdef}
\end{equation}
where $n(M)\mathrm{d}M$ is the comoving number density of halos with mass between $M$ and $M+\mathrm{d}M$, and $W(k,M)$ is the Fourier transform of the halo density profile normalized by the matter density at $z=0$, $\bar{\rho} \propto \omega_m$. The quantities on the right-hand side of Equation~\eqref{eq:1hdef} are evaluated at the relevant redshift. We have
\begin{equation}
  W(k,M) = 4\pi \int_0^\infty r^2 \mathrm{d}r \frac{\sin(kr)}{kr} \frac{\rho(M,r)}{\bar{\rho}}.
\end{equation}

We assume that the halo mass function can be written as a universal function of $\nu \equiv \delta_c/\sigma(M)$, where $\delta_c \approx 1.686$ is the spherical collapse threshold and $\sigma^2(M)$ is the linear variance in spheres of mass $M$, where $M/\bar{\rho} =  4\pi R^3/3$ and
\begin{equation}
  \sigma^2(R) = \int_0^\infty \frac{\mathrm{d}k}{k} \Delta_L^2(k) \left[\frac{3}{(kR)^3}(\sin kR - kR \cos kR)\right]^2.
\end{equation}

Defining the function $f(\nu)\mathrm{d}\nu \equiv (M/\bar{\rho})n(M)\mathrm{d}M$, the large-scale limit of the 1-halo term can be written as
\begin{equation}
  \lim_{k \to 0} \Delta_{1H}^2(k) = \frac{(k/h)^3}{2\pi^2} \frac{h^3}{\bar{\rho}} \int_0^\infty M(\nu) f(\nu) \mathrm{d}\nu,
  \label{eq:1hshot}
\end{equation}
where $M(\nu)$ is the inverse of $\nu(M)$. Note that $f(\nu)$ integrates to unity over the range of its argument, since all matter is assumed to be in halos.
\begin{figure}
  \includegraphics[width=\columnwidth]{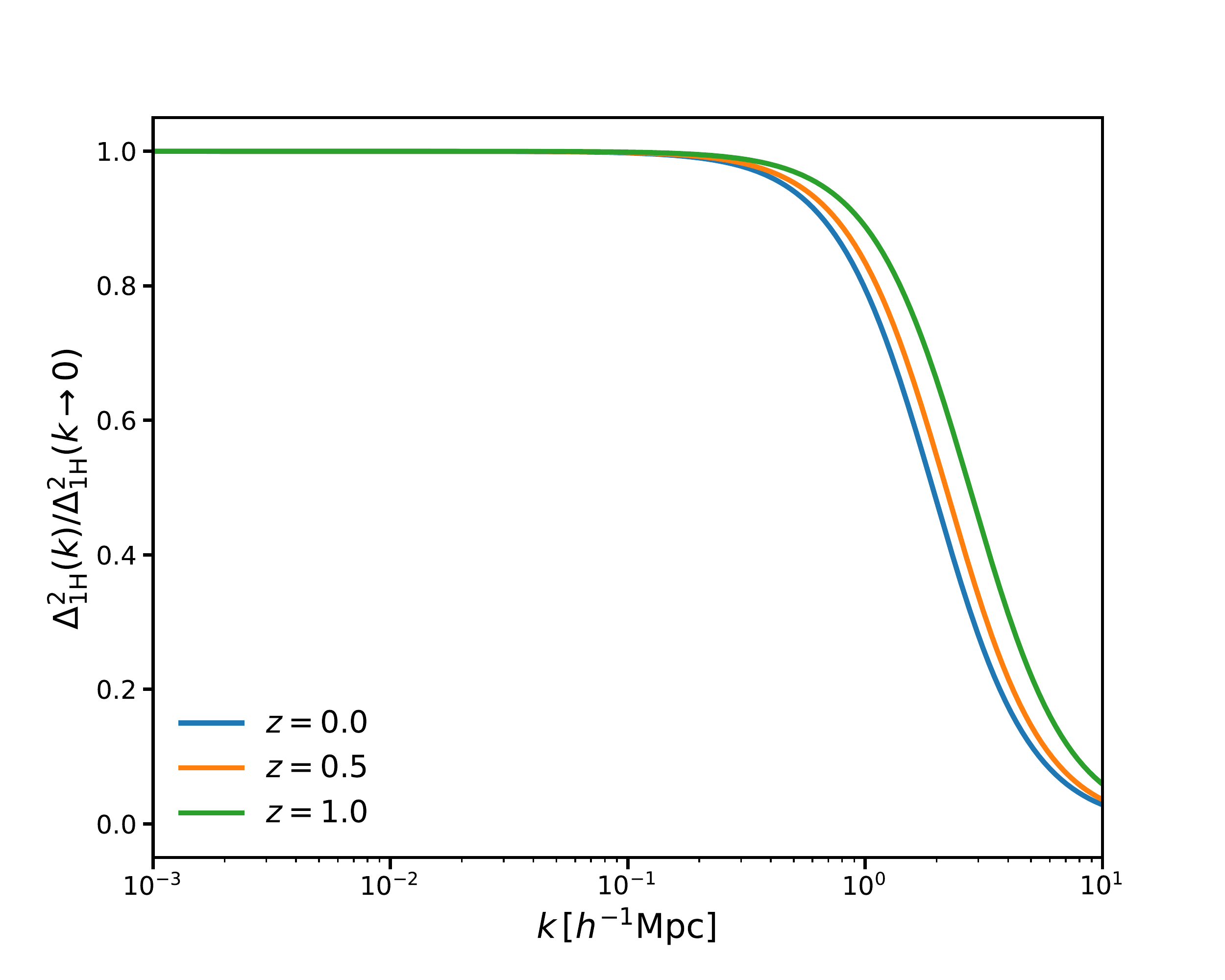}
  \caption{The scale dependence of the 1-halo term normalised by its large-scale amplitude at $z=0$ (blue), $z=0.5$ (orange) and $z=1$ (green).}
  \label{fig:PK_1H_fid}
\end{figure}

The amplitude of the 1-halo term is mostly determined by its $k \to 0$ limit where it tends to a constant shot noise term. On scales comparable to the halo size the 1-halo term is suppressed relative to the extrapolation of its large-scale amplitude, and the details of the halo density profile become important. This behaviour is shown in Figure~\ref{fig:PK_1H_fid} for a few relevant redshifts. Following the fiducial KV450 analysis and the \textsc{hmcode} prescription, we assume halos have the density profile of~\citet[][hereafter NFW]{1997ApJ...490..493N}, truncated at a radius $r_v$ where the enclosed mass is a fraction $\Delta_v$ of the $z=0$ background. In \textsc{hmcode}, $\Delta_v$ is given redshift dependence and is purely a function of $\Omega_m$ in $\Lambda$CDM models. The scales probed by lensing are mostly sensitive to the $k \to 0$ limit of the 1-halo term and the transition region where the density profile becomes important. The cosmology dependence of the former can be studied by examining Equation~\eqref{eq:1hshot}, where the details of the halo profile are irrelevant.

Ignoring the weak cosmology dependence of $\delta_c$, the cosmology dependence of the 1-halo amplitude follows from that of $M(\nu)$ integrated against $f(\nu)$. As pointed out earlier, the slope of $\sigma(R)$ gets flatter at lower $R$, meaning $M(\nu)$ is increasingly sensitive to $\nu$ at small values of $\nu$, and asymptotes at large $\nu$ to $M(\nu) \propto \nu^{3/2}$. Although the halo mass function is peaked around $\nu \approx 1$~\citep{1999MNRAS.308..119S}, a steep $M(\nu)$ can push the relevant $\nu$ scales to higher values resulting in a change to the cosmology dependence of the integral. In general therefore it is necessary to simultaneously determine the relevant values of $\nu$ that contribute to the integral in Equation~\eqref{eq:1hshot} and the slope of $M(\nu)$. At $z=0$ we find the most relevant scales are $R \sim 10 \, h^{-1} \, \mathrm{Mpc}$ for which $\nu \approx 2$, where the variance in spheres can be approximated (assuming $n_s = 1$) in the model of Equation~\eqref{eq:D2lin} as
\begin{equation}
  \sigma^2(R) \propto A_s \frac{k_{\mathrm{eq}}^4}{\omega_m^2} g^2(a=1) \; \omega_m^a \; (k_{\mathrm{eq}}R)^b,
\end{equation}
Note that when $R = 8 \, h^{-1} \, \mathrm{Mpc}$ we have $a \approx 0.45$ and $b \approx -1.4$, which gives the cosmology scaling of $\sigma_8$ in terms of $A_s$. Plugging this into the definition of $\nu$ gives
\begin{align}
  M(\nu) & \propto \, A_s^{-3/b}  \omega_m^{1-3(2+a+b)/b} [g(a=1)]^{-6/b} \nu^{-6/b} \nonumber \\
  &  \propto \, \omega_m \sigma_8^{-6/b} \Omega_m^{-(3a + 3b + 2.85)/b} h^{-(6a + 6b + 1.5)/b} \nu^{-6/b}.
  \label{eq:Mnu}
\end{align}
Using the scaling found for $\sigma_8$ implies that
\begin{equation}
  M(\nu)/\bar{\rho} \sim \sigma_8^{4.3} h^{-3} \nu^{4.3}
\end{equation}
around $\nu \approx 2$, with no dependence on $\Omega_m$. The parameter dependence of the integral in Equation~\eqref{eq:1hshot} can thus be approximated as
\begin{equation}
  \lim_{k \to 0} \Delta_{1H}^2(k) \propto (k/h)^3 \sigma_8^{4.3}.
  \label{eq:1hamp}
\end{equation}
Remarkably, both $h$ and $\Omega_m$ have completely dropped out of the parameter dependence of the 1-halo amplitude in this approximation. This is a result of normalizing by $\sigma_8$ rather than $A_s$\footnote{This is a good reason for using $\sigma_8$ rather than the $\sigma_{12}$ parameter advocated by~\citet{2020PhRvD.102l3511S}.} and the fact that most of the contribution to the 1-halo integral comes from scales around $R \approx 8 \, h^{-1} \, \mathrm{Mpc}$. The latter is a consequence of $\sigma_8$ being of order unity at $z=0$ (indeed, this is why the parameter was originally introduced into cosmology). In practice the 1-halo integral is not purely sensitive to the slope of $M(\nu)$ around $\nu \approx 2$ and range of $\nu$ contribute, which leads to slightly different values of $a$ and $b$ in Equation~\eqref{eq:Mnu} and some non-zero dependence of the amplitude on $\Omega_m$ and $h$; we find $b\approx -1.3$ and $a \approx 0.3$ give a slightly better fit to the parameter dependence.
  
Reinstating the halo density profile, the full 1-halo term can be written as
\begin{equation}
  \Delta_{1H}^2(k) = \frac{k^3}{2 \pi^2} \frac{1}{\bar{\rho}} \int_0^\infty \mathrm{d}\nu \, f(\nu) M(\nu) F^2(k r_s),
\end{equation}
where $F(k r_s) = W(k,M)/W(0,M)$ and $r_s$ is the scale radius of the NFW profile, whose cosmology dependence is given by
\begin{equation}
  r_s \propto M^{1/3} \Delta_v^{-1/3} \bar{\rho}^{-1/3}c^{-1}(M, z)
\end{equation}
where $c(M,z)$ is the halo concentration parameter. Neglecting the weak $M$ dependence of $c$, taking $\Delta_v \propto \Omega_m^\gamma$, and using Equation~\eqref{eq:Mnu} at $z=0$ gives locally
\begin{align}
  k_s \equiv r_s^{-1} & \propto A_s^{1/b} \omega_m^{(2+a+b)/b} [g(a=1)]^{2/b} \Omega_m^{\gamma/3}\nu^{2/b} \nonumber \\
  & \propto \sigma_8^{2/b} \Omega_m^{(0.95+a+b)/b + \gamma/3} h^{(0.5 +2a + 2b)/b} \nu^{2/b}.
\end{align}
Taking the default value of $\gamma = -0.35$ used in~\citep{2015MNRAS.454.1958M} and assuming the same values of $a$ and $b$ as used in the low-$k$ limit, we find
\begin{equation}
  k_s/h \sim \sigma_8^{-1.4} \Omega_m^{0.1}.
  \label{eq:1hks}
\end{equation}
The independence from $h$ in this expression reflects the fact that the two scales $r_s$ and $r_v$ both scale with the effective halo volume $(M/\bar{\rho})^{1/3}$ in this model, and as discussed above the normalization by $\sigma_8$ absorbs almost all the $h$-dependence of the $M(\nu)$ relation. As for the 1-halo amplitude, a range of $\nu$ values contribute to $k_s$, which imparts additional $\Omega_m$ and $h$ dependence to $k_s$.

Since the density profile part of the 1-halo term suppresses power (see Figure~\ref{fig:PK_1H_fid}), increasing $\sigma_8$ decreases $k_s$ and increases this suppression, opposing the increase in power favoured by the amplitude increase seen in Equation~\eqref{eq:1hamp}. This results in a `bump' in the $\sigma_8$-dependence of the total matter power spectrum; as $k$ increases from linear the increasingly important 1-halo amplitude enhances sensitivity through its steep $\sigma_8$ scaling. As $k$ is increased further this sensitivity is suppressed by the density profile part of the 1-halo term, ultimately causing its sign to change at very high $k$. The bump is clearly visible around $k \approx 1 \, h \, \mathrm{Mpc}^{-1}$ at $z=0$ in the halo model response plots of~\citet{2015MNRAS.454.1958M, 2020MNRAS.493.1640C} and can also be seen in the N-body responses of~\citet{2014ApJ...780..111H}.

Putting these results together, we can see that when expressed as a function of $k/h$, the 1-halo power spectrum at $z=0$ is very insensitive to changing $h$ at fixed $\sigma_8$ and $\Omega_m$. Moving to higher redshifts, the growth factor $g(a)$ tends to unity which changes the range of $\nu$ contributing most to the 1-halo integral and imparts additional cosmology dependence, although the effect is modest.

This suggests that the lensing power spectrum on 1-halo scales will be very insensitive to $h$ when expressed in terms of $\sigma_8$ and $\Omega_m$. From Equation~\eqref{eq:Clgg_scaled} we see that the 1-halo lensing power spectrum should scale roughly as $C_\ell \sim \sigma_8^{4}\Omega_m^2$, i.e. roughly as $S_8^4$. This recovers the canonical $S_8$ scaling in agreement with the results of~\citetalias{1997ApJ...484..560J}, but now using an independent and more accurate model for the non-linear matter power spectrum.

The cosmology dependence of the total lensing power spectrum depends on the relative contributions of the 1-halo and 2-halo terms. We have shown that on large linear scales we expect $C_\ell \sim \sigma_8^2 \Omega_m^{-1} h^{-3.5} \sim S_8^2\omega_m^{-2}h^{0.5}$, transitioning first to $C_\ell \sim \sigma_8^2 \Omega_m^{3.7} h^{1.9} \sim S_8^2 \omega_m^{2.7} h^{-3.5} $ and then to $C_\ell \sim \sigma_8^{4}\Omega_m^2 \sim S_8^4$ as the 1-halo term becomes important. We expect the $\sigma_8$ dependence to decrease dramatically on smaller angular scales that are sensitive to the halo density profile, with small additional $h$ and $\Omega_m$ dependence entering. On scales where current lensing surveys have most signal-to-noise the amplitude of $C_\ell$ scales as $S_8^\alpha$ with $2 < \alpha < 4$. Therefore, information on $\Omega_m$ and $h$ separately can only come from precise measurements of both large linear scales and small very non-linear scales.

We close this section by noting that our discussion so far has focussed on shear two-point correlation functions and angular power spectra. Several alternative two-point estimators have been successfully applied to weak lensing data, for example, the real-space aperture mass statistics and the COSEBI statistics. The parameter information brought by a given estimator ultimately depends on which physical scales receive highest weight in the likelihood function after scale cuts have been applied, as well as the effects of parameter degeneracies - both these effects depend on the estimator in question. Specialising our discussion to the angular power spectrum is advantageous since many alternative estimators may be written as a linear transform of the angular power spectrum. In addition, the dominant (Gaussian) part of the covariance matrix is diagonal for the angular power spectrum, which facilitates a study of parameter information across angular scales.

\subsection{Numerical results}
\label{subsec:numerical}

\subsubsection{Matter power spectrum}
\label{subsubsec:Pknumerical}


\begin{figure*}
  \includegraphics[width=\textwidth]{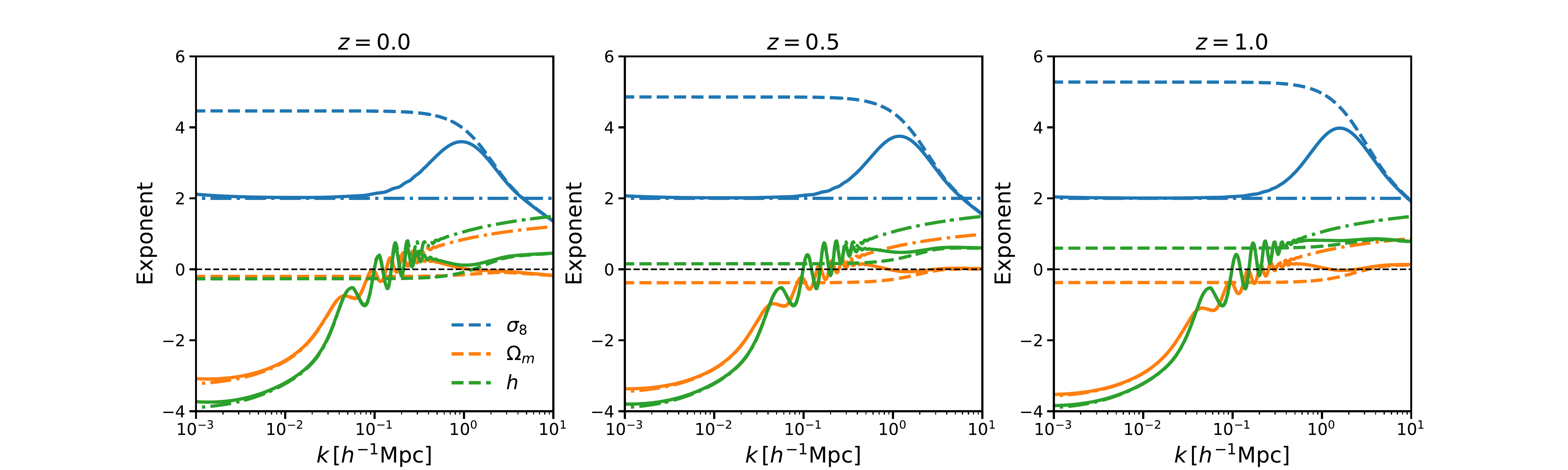}
  \caption{The parameter dependence of the matter power spectrum in linear theory (dot-dashed), the 1-halo term (dashed) and the total (solid) for the three parameters $\sigma_8$ (blue), $\Omega_m$ (orange) and $h$ (green) at $z=0$ (left panel), $z=0.5$ (middle panel) and $z=1$ (right panel). Plotted are the exponents of a power law dependence around the fiducial model, i.e. $\Delta^2(k/h) \propto \sigma_8^\alpha \Omega_m^{\beta} h^\gamma$. The derivatives were computed using the halo model. Flat behaviour over a range of $k$ indicates that only the amplitude is changing with the parameter, whereas linear behaviour indicates that the slope is also changing.}
  \label{fig:PK_individual_dependencies}
\end{figure*}

In Figure~\ref{fig:PK_individual_dependencies} we plot the parameter dependence of the matter power spectrum as a function of $k/h$, for a few relevant redshifts in linear theory and in the halo model\footnote{We use an implementation of the halo model packaged with~\href{https://github.com/alexander-mead/HMcode}{\textsc{hmcode}}.}. Our toy model for the linear power spectrum Equation~\eqref{eq:linear_model} works well on all scales. On large linear scales at $z=0$ the dependence on parameters is roughly $\Delta^2 \propto \sigma_8^2 \Omega_m^{-3} h^{-3.7}$, which agrees well with the prediction of Equation~\eqref{eq:D2lin}, and on small scales the dependence of the linear power on $\Omega_m$ and $h$ asymptotes to values consistent with Equation~\eqref{eq:D2lin}. Note that we do not attempt to model the detailed cosmology dependence of the BAO feature, as this will be mostly washed out in the lensing projection.

The parameter dependence of the total power spectrum on non-linear scales is determined by that of the 1-halo term. Figure~\ref{fig:PK_individual_dependencies} confirms our analytical argument that the cosmology dependence of the 1-halo term at $z=0$ is almost entirely captured by $\sigma_8$ (see Equations~\eqref{eq:1hamp} and~\eqref{eq:1hks}). As expected from the arguments above, the dependence of $\Delta^2$ on $\sigma_8$ is quadratic on large scales and displays a `bump' on small scales as the 1-halo comes to dominate the power. Figure~\ref{fig:PK_individual_dependencies} clearly displays the falling sensitivity of the 1-halo term to $\sigma_8$ on scales where the density profile is important.

The power spectrum at $z>0$ is generally more sensitive to all parameters compared with at $z=0$. In particular the $\sigma_8$ dependence of the 1-halo amplitude increases due to a decrease in the Lagrangian length scales contributing to the halo shot noise. Smaller Lagrangian patches where the local variance is higher are more likely to collapse relative to large patches, and this is true to a greater extent at high redshift than at $z=0$ due to the lower amplitude of fluctuations on all scales (`hierarchical growth'). The slope of $\sigma(R)$ is flatter at lower $R$, so $R(\sigma)$ is steeper at higher $\sigma$ (and hence lower $R$). Since the amplitude of the halo shot noise scales as $R^3$, this boosts the $\sigma_8$ sensitivity at high redshift, and since the scales that contribute most are now less than $8 \, h^{-1} \, \mathrm{Mpc}$ additional dependence on $\Omega_m$ and $h$ arises. Despite this, the dependence on non-$\sigma_8$ parameters remains weak (sub-linear) on all scales in the 1-halo term out to $z=1$.


\subsubsection{Lensing power spectrum}
\label{subsubsec:lensingnumerical}

\begin{figure*}
  \includegraphics[width=\textwidth]{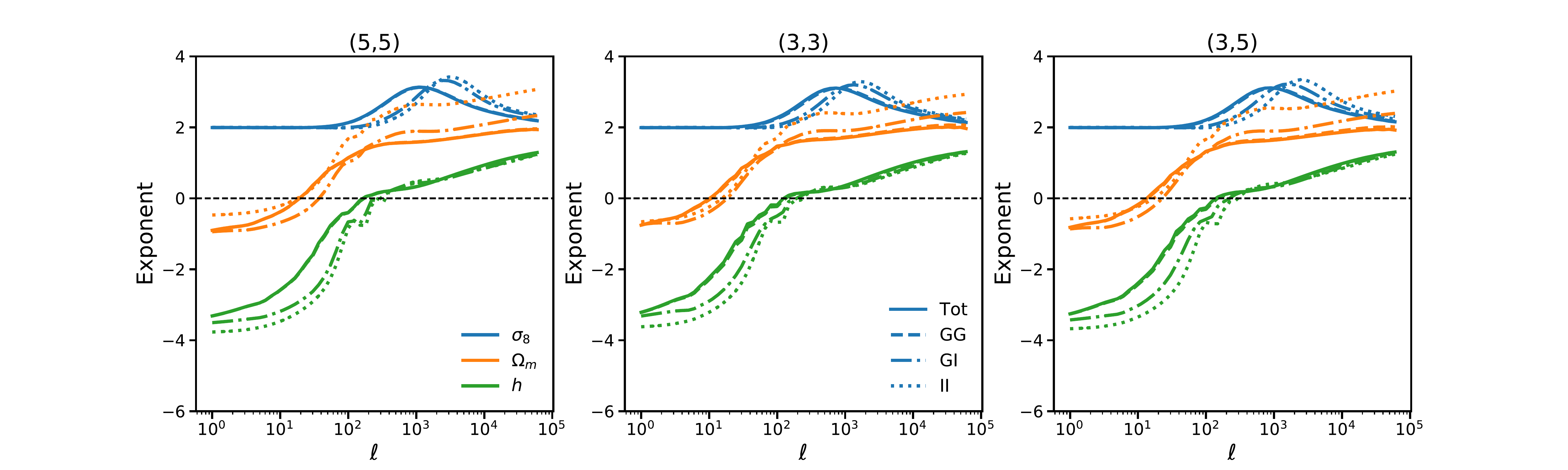}
  \caption{The parameter dependence of the lensing angular power spectrum, showing the individual contributions from the GG term (dashed), the GI term (dot-dashed), the II term (dotted), and the total (solid). We show the autospectrum for KV450 redshift bin 5 (central redshift $z=1$, left panel), the autospectrum of redshift bin 3 (central redshift $z=0.6$, middle panel), and the cross-spectrum of these bins (right panel). Plotted are the exponents of a power law dependence around the fiducial model, i.e. $C_\ell \propto \sigma_8^\alpha \Omega_m^{\beta} h^\gamma$.}
  \label{fig:Cl_KV450_dependcies}
\end{figure*}

In Figure~\ref{fig:Cl_KV450_dependcies} we show the dependence of the lensing power spectrum (both from shear and from intrinsic alignments) on parameters for a few different KV450 redshift bins. The detailed dependence at each $\ell$ follows from the mapping $k$ to $\ell$ determined by the Limber approximation, which is shown in Figure D1 of~\citet{2020A&A...641A.130M}. Roughly speaking, most of the signal to noise in KV450 is from $10^2 \lesssim \ell \lesssim 10^4$, which corresponds to $0.1 \lesssim k / h \, \mathrm{Mpc}^{-1} \lesssim 10$ across all redshift bins.

The total lensing power spectrum is dominated by the shear-shear power spectrum for these redshift bin combinations, but we note that the cosmology dependence of the IA amplitude is not drastically different due in part to the fact that in the linear alignment model~\citep{2004PhRvD..70f3526H} IAs are proportional to the local gravitational tidal field, just as in lensing.

As in the case of the matter power spectrum the dependence on $\sigma_8$ exhibits a `bump' around $\ell \sim 10^3$, arising from the interplay of the 1-halo and 2-halo terms (now in projection). On large linear scales the dependence is roughly $C_\ell \propto \sigma_8^2 \Omega_m^{-1} h^{-3.5}$, recovering our analytic prediction almost perfectly. The dependence on $\Omega_m$ and $h$ increases with $\ell$, again in agreement with our linear expectation. On scales where the 1-halo term dominates the $h$ dependence is very weak (sub-linear), remaining so out to $\ell \sim 10^4$. Across these same scales the power spectrum depends on $\Omega_m$ as roughly $\Omega_m^{1.5}$, which is comparable to our prediction based purely on the 1-halo amplitude of $\Omega_m^2$.

The net effect of these parameter dependencies is a scaling of $C_\ell \propto \sigma_8^2 \Omega_m$ around $\ell \sim 10^2$, rising to $C_\ell \propto \sigma_8^3 \Omega_m^{1.5}$ at $\ell \sim 10^3$, before falling to $C_\ell \propto \sigma_8^{2.5} \Omega_m^{1.5}h^{0.5}$ at $\ell \sim 10^4$. Importantly, across most of the angular scales where KV450 (and other current cosmic shear surveys) have high signal to noise, the lensing power spectrum depends primarily on the combination $S_8$ with almost no additional dependence on $h$. On the smallest angular scales measured well this degeneracy is slightly broken, with the dependence more like $S_8^{2.5} \omega_m^{0.25}$. These small scales are highly influenced by baryon feedback so the implications for posteriors are unclear, but this may be the origin of the weak constraint on $\omega_m$ seen in Figure~\ref{fig:KV450_Kns_BBN_cosmoparams}.

We have so far focussed on the scaling of the power spectrum per-$\ell$, but we can also consider the parameter dependence of the various power-law slopes and amplitudes present in the spectrum. This information is partly contained in the logarithmic derivatives plotted in Figure~\ref{fig:Cl_KV450_dependcies}. These curves show the quantities $\alpha$, $\beta$, and $\gamma$, where $C_\ell \propto \sigma_8^\alpha \Omega_m^{\beta} h^\gamma$. If these curves are constant over some range of $\ell$ this implies that only the amplitude of $C_\ell$ is being changed across this $\ell$-range. If the curves are linear in $\ell$ with zero intercept this implies that the slope is changing at fixed amplitude. Figure~\ref{fig:Cl_KV450_dependcies} shows that, at fixed $\sigma_8$ and $\Omega_m$, changing $H_0$ changes the slope of the power spectrum on scales $10^3 \lesssim \ell \lesssim 10^4$.

Break scales also contribute information on $H_0$. The lensing power spectrum essentially contains three angular scales: the equality scale of matter-radiation equality $\ell_{\mathrm{eq}}$, the scale where the power spectrum transitions between the 2-halo and 1-halo regimes $\ell_{{\rm NL}}$, and the angular scale associated with the NFW scale radius, which we can define as $\ell_s = \chi_*/r_s$ for a source at comoving distance $\chi_*$. The scales $\ell_{{\rm NL}}$ and $\ell_s$ are the most well placed to be measured by current surveys, but their cosmological information content is contaminated by uncertainties in the baryon feedback model. We find that $\ell_{{\rm NL}} \sim \sigma_8^{-0.8}\Omega_m^{0.4}h^{0.5}n_sB_{{\rm bary}}$, where $B_{{\rm bary}}$ is the baryon feedback amplitude. Furthermore, $\ell_{{\rm NL}}$ is not a sharp break in the power spectrum but more a broad transition region between linear and non-linear scales, so carries little useful $H_0$ information. As shown in Equation~\eqref{eq:1hks}, $\ell_s$ depends on parameters in the standard halo model roughly as $\sigma_8^{-1.4}\Omega_m^{0.1}$, which might raise hopes that one can measure $\sigma_8$ and $\Omega_m$ separately by measuring this scale. However the \textsc{hmcode} prescription for baryon feedback adds extra dependence on $B_{{\rm bary}}$, which complicates this argument. If neutrino mass is also marginalised over we can expect the usefulness of these scales in constraining $H_0$ to be diluted further. $\ell_{\mathrm{eq}}$ is in principle cleaner but is not well measured in current lensing surveys. This scale should however be accessible to forthcoming wide surveys aiming to measure shear over large fractions of the sky\footnote{Note that there is likely to be some information on $\ell_{\mathrm{eq}}$ in current lensing surveys; Figure~\ref{fig:KV450_xi_kernels} demonstrates that $\ell < 100$ contributes non-negligibly to the two largest scale $\xi_+$ measurements and the largest scale $\xi_-$ measurement. This information could be contributing to the weak $\omega_m$ constraint seen in Figure~\ref{fig:KV450_Kns_BBN_cosmoparams}.}. A measurement of this scale would provide information on the combination $\Omega_m^\alpha h$ with $\alpha < 1$.

To a reasonable approximation current lensing surveys measure a range of scales where the matter power spectrum can be modelled as a pure power law. There is also information on $\Lambda$CDM parameters from the redshift \emph{evolution} of the power spectra, but we found this to be insignificant in the case of $\sigma_8$, $\Omega_m$, and $h$\footnote{Obviously the redshift dependence will be crucial in measuring non-standard expansion histories, such as those arising in dynamical dark energy models.}.

\begin{figure}
  \includegraphics[width=\columnwidth]{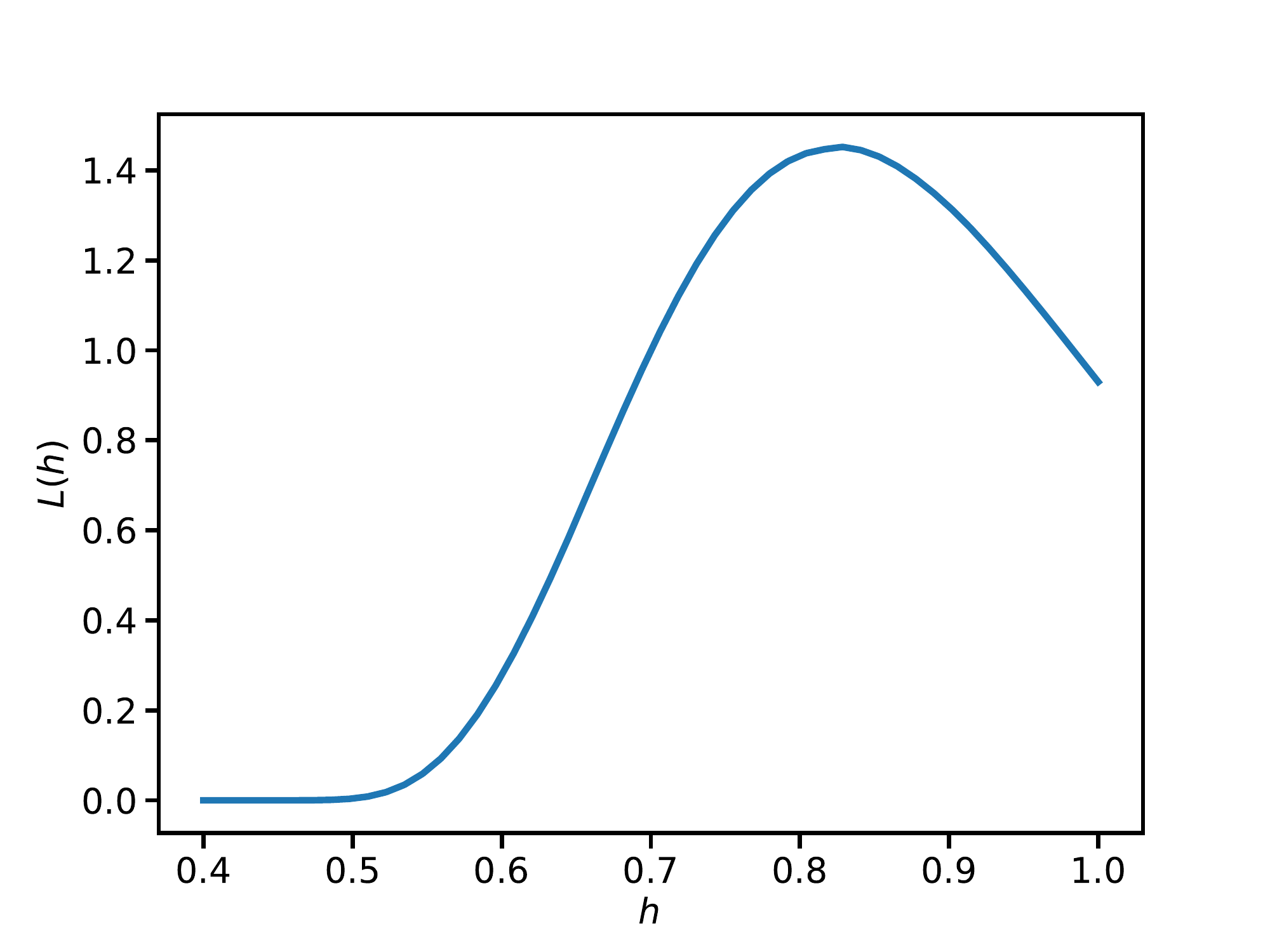}
  \includegraphics[width=\columnwidth]{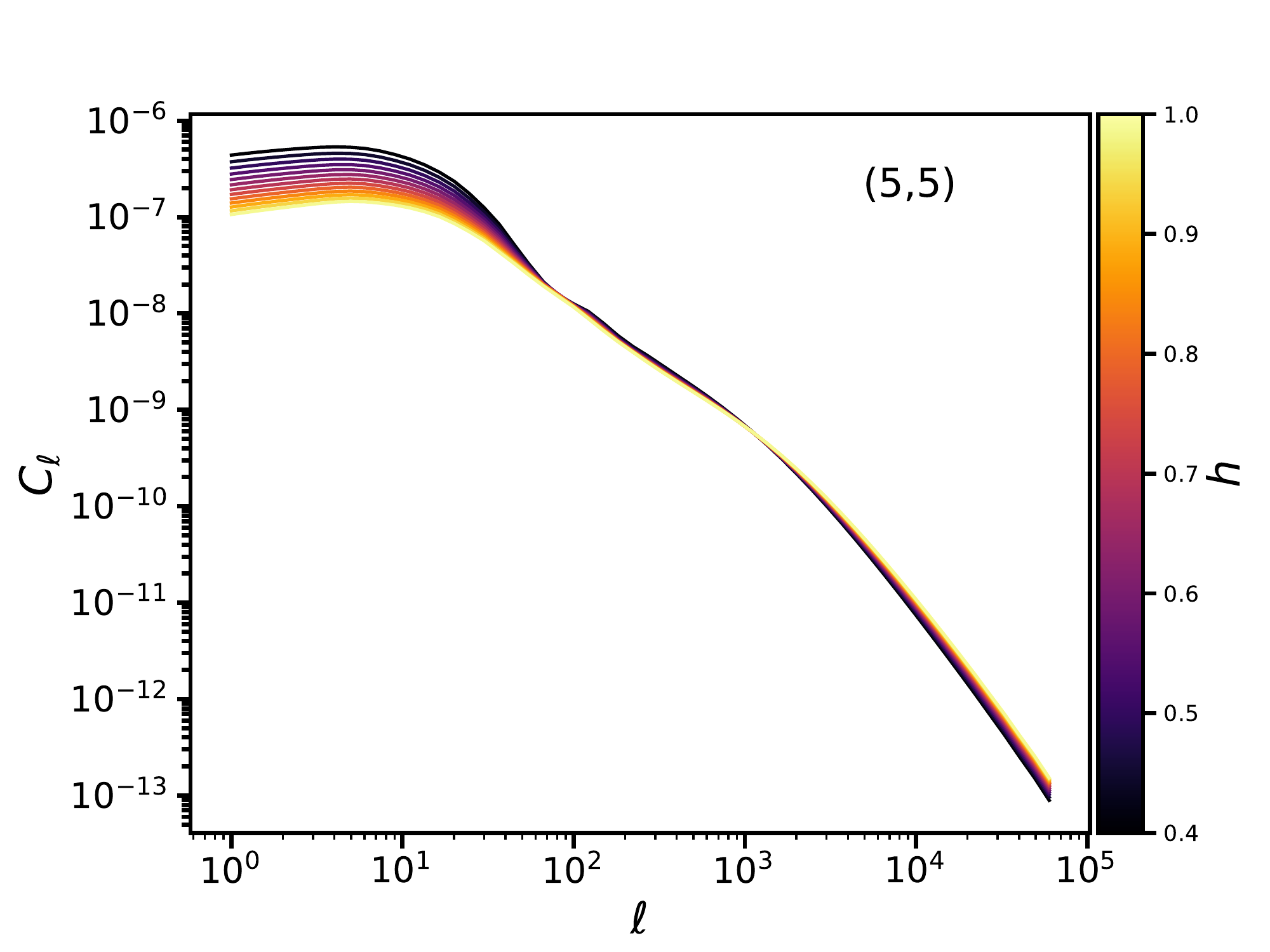}
  \caption{\emph{Top panel}: The KV450 likelihood as $h$ is varied, fixing all other parameters (including $\Omega_m$ and $\sigma_8$), with arbitrary normalisation. \emph{Bottom panel}: The corresponding change to the lensing angular power spectrum for the highest redshift bin autospectrum. Note that the normalization of the likelihood function is arbitrary here.}
  \label{fig:KV450_varying_h_fixed_S8_Omm}
\end{figure}

In Figure~\ref{fig:KV450_varying_h_fixed_S8_Omm} we explicitly show how the low sensitivity to $h$ across well-measured scales in KV450 manifests in the likelihood function. We show the change in the lensing power spectrum as $h$ is uniformly varied within its prior range while fixing $\sigma_8$, $\Omega_m$ (and hence $S_8$), and all other parameters, along with the corresponding change to the likelihood. The stark insensitivity to $h$ in the region $10^2 \lesssim \ell \lesssim 10^3$ is clearly visible. The preference for values of $h \approx 0.8$ is mostly explained by the well-measured small scales where residual $h$ effects show up (although the sensitivity is suppressed by non-diagonal terms in the covariance matrix that show up on these scales) and large scales where the sensitivity is higher but the contribution to the signal is lower.

\begin{figure}
  \includegraphics[width=\columnwidth]{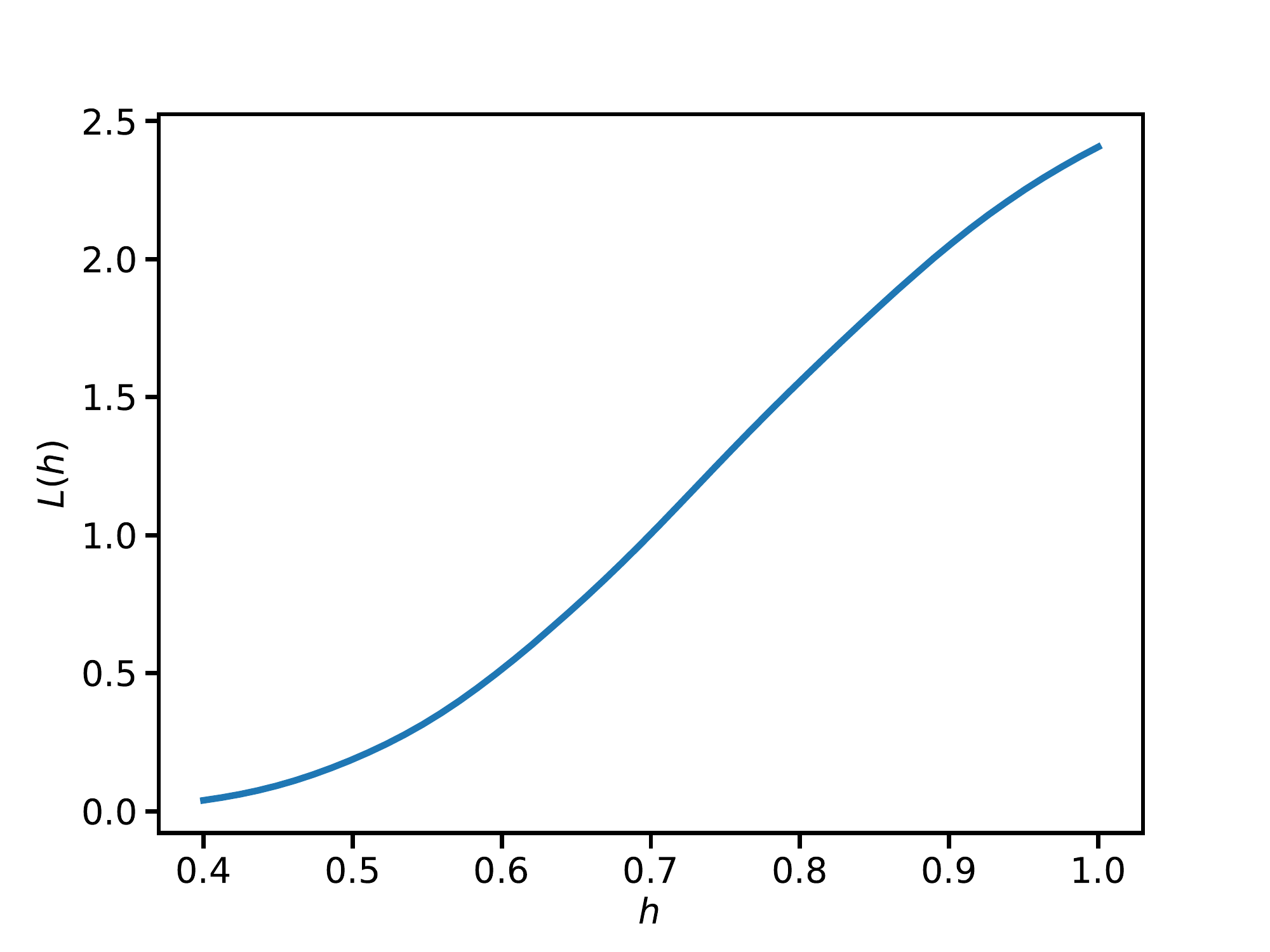}
  \includegraphics[width=\columnwidth]{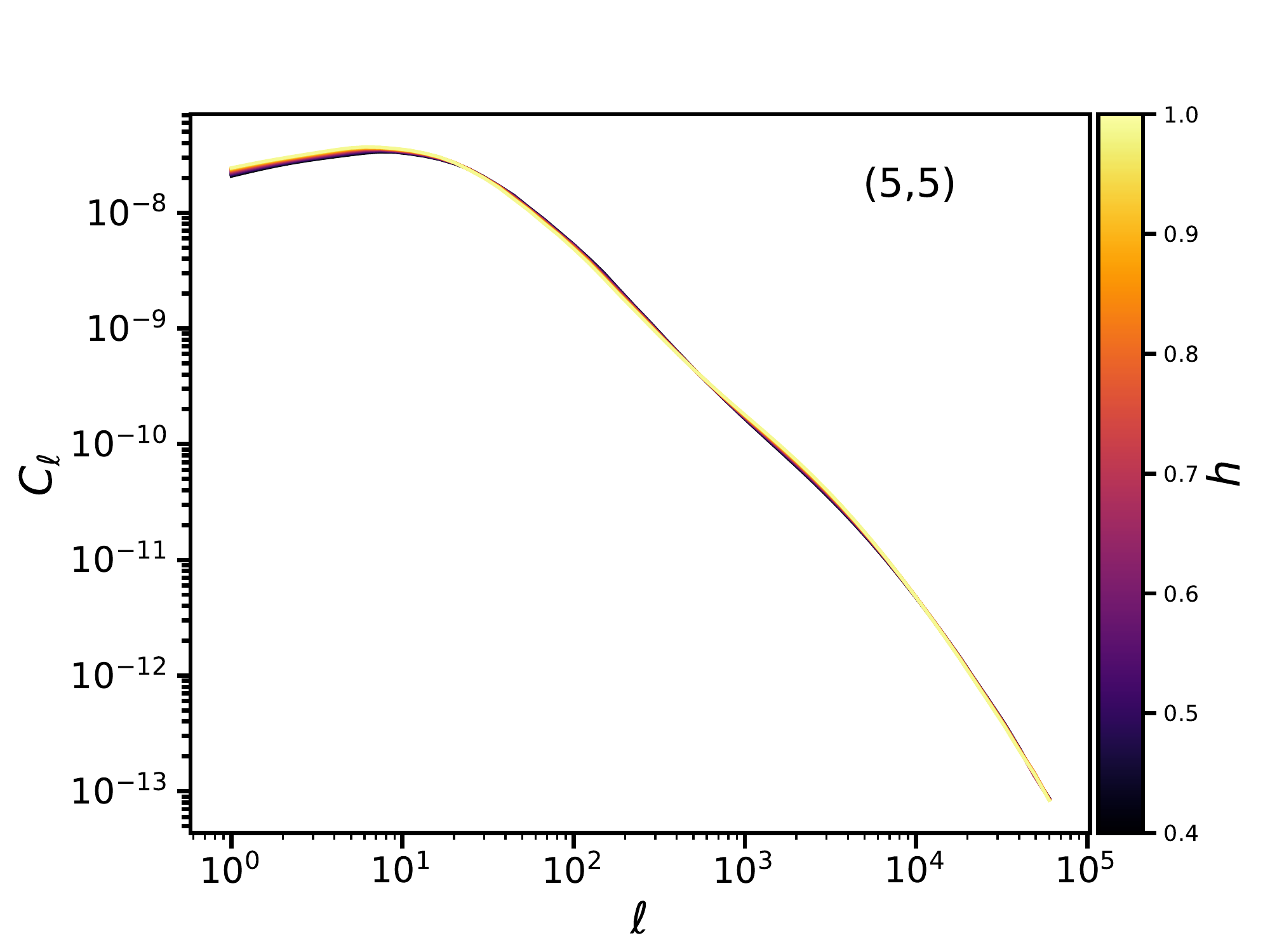}
  \caption{Same as Figure~\ref{fig:KV450_varying_h_fixed_S8_Omm} but fixing $\Omega_m h^2$ and $S_8$, changing $\sigma_8$ and $\Omega_m$ to compensate. Note that the normalization of the likelihood function is arbitrary here.}
  \label{fig:KV450_varying_h_fixed_S8_ommh2}
\end{figure}

In Figure~\ref{fig:KV450_varying_h_fixed_S8_ommh2} we show how the likelihood and power spectrum changes when $h$ is varied but $\omega_m$ is kept fixed (along with $S_8$, hence $\sigma_8$ and $\Omega_m$ are changed to compensate). The likelihood peak has gone, but there is still preference for high $h$. The small residual changes to the $C_\ell$ across this range of $h$, now barely visible in Figure~\ref{fig:KV450_varying_h_fixed_S8_ommh2}, actually sum coherently to give $\sim 10\%$ changes in the correlation functions on angular scales $10' \lesssim \theta \lesssim 100'$. Some of these scales are measured with signal-to-noise of roughly a few in some of the higher redshift bin combinations. It is therefore possible that the weak $h$ constraint in KV450 comes from sensitivity to linear scales $50 \lesssim \ell \lesssim 100$ (based on Figure~\ref{fig:KV450_xi_kernels}), in particular from changes to the equality scale $\ell_{{\rm eq}} \propto \omega_m/h$ at fixed $\omega_m$ that keep the small-scale shape roughly the same. The constraint is very weak, and potentially influenced by the hard priors on other parameters, so we choose not to investigate it further.
%


\section{Future prospects}
\label{sec:future}

\begin{figure}
  \includegraphics[width=\columnwidth]{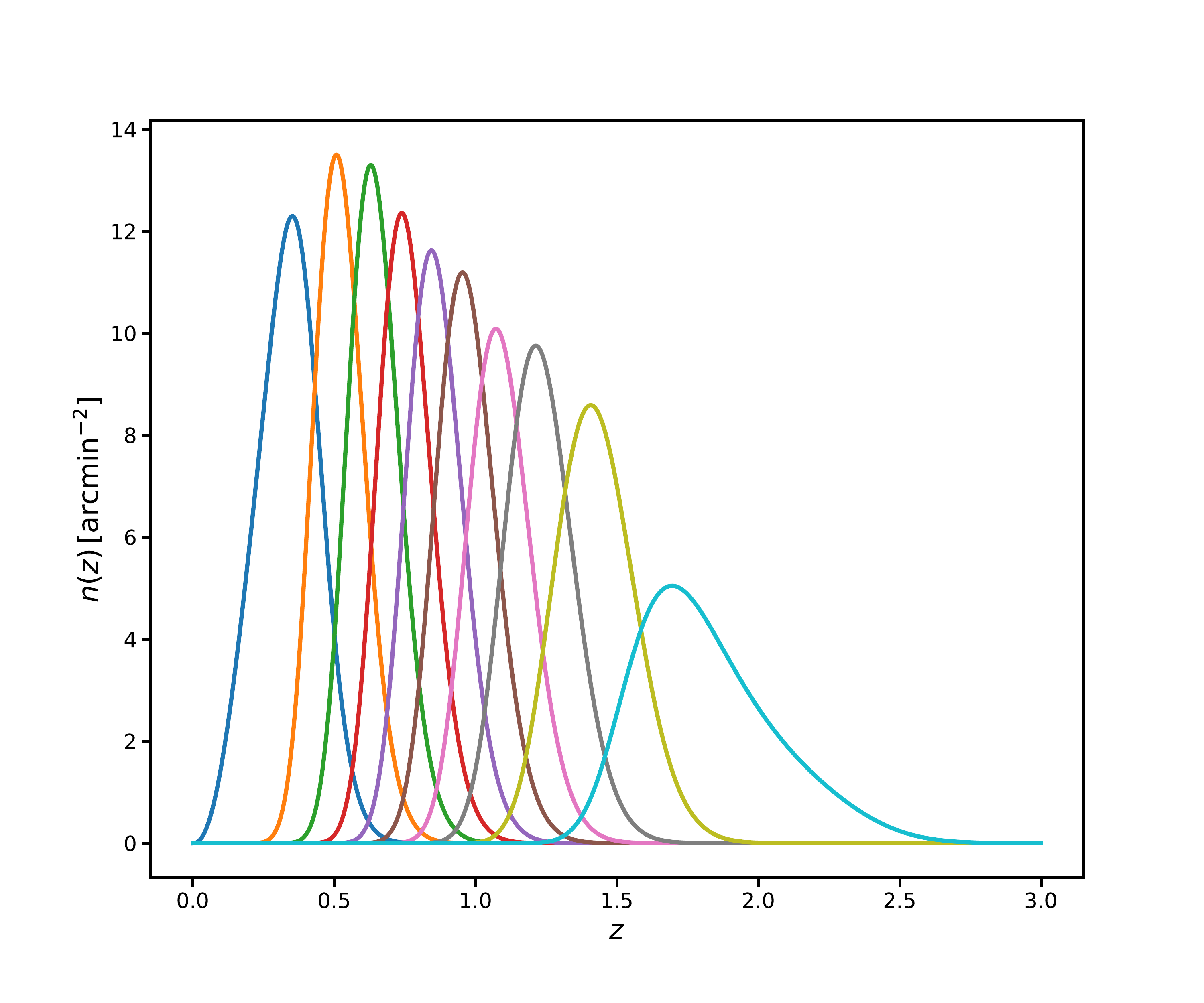}
  \caption{Fiducial set of redshift bins adopted for our Euclid-like lensing forecast. Specifications have been taken from~\citealt{2020A&A...642A.191E}.}
  \label{fig:Euclid_nzs}
\end{figure}

We have seen that current cosmic shear experiments are not able to constrain $H_0$ without external data. In this section we investigate whether this will be the case for forthcoming Stage-IV lensing surveys. We consider a future Euclid-like weak lensing survey mapping cosmic shear across 15,000 sq. deg. with a source number density of $\bar{n} = 30 \, \mathrm{arcmin}^{-2}$. We assume that photometric redshifts are available such that source galaxies can be placed in one of ten redshift bins, which we define to be equipopulated and broadened due to photometric redshift errors as modelled in~\citet{2020A&A...642A.191E}. These redshift bins are shown in Figure~\ref{fig:Euclid_nzs}.

We compute the Fisher matrix assuming a diagonal Gaussian covariance for angular power spectrum estimates across a range of angular multipoles from $\ell_{\mathrm{min}} = 10$ to $\ell_{\mathrm{max}} = 5000$. The cosmic variance part of this covariance is fixed to a fiducial $\Lambda$CDM model with parameters given by the best-fitting Planck 2018 model~\citep{2020A&A...641A...6P} with a total neutrino mass fixed to 0.06eV assuming one massive and two massless neutrino species. For the computations in this section we use \textsc{camb} (v1.3.0, \citealt{2000ApJ...538..473L, 2012JCAP...04..027H}) to compute the linear matter power spectrum, with baryon feedback modelled with the one-parameter model of~\citet{2021MNRAS.502.1401M} included in the latest \textsc{hmcode}. Shape noise is included in the Fisher matrix assuming an ellipticity standard deviation of $0.21$ per component. Derivatives of the power spectrum are computed at the same fiducial model as the covariance matrix. Intrinsic alignments are modelled with the `non-linear linear alignment' model described in Section~\ref{sec:WLcosmodep}, modelled with a free amplitude parameter $A_{IA}$. We also vary the amplitude of baryon feedback using the parameter $\log_{10} T_{\mathrm{AGN}}/\mathrm{K}$ described in~\citet{2021MNRAS.502.1401M}. Our Fisher matrix differs slightly to that of~\citet{2020A&A...642A.191E, 2020MNRAS.493.1640C} due to our use of the more recent \textsc{hmcode} as well as our implicit assumption that an E/B mode decomposition can be made such that only the per-component shape noise variance contributes to the data vector. We account for the loss of modes due to the sky mask with an $f_{\mathrm{sky}}$ factor in the Fisher matrix. Note that our constraints are likely over-optimistic given our use of a diagonal covariance matrix\footnote{The dominant non-Gaussian contribution to the covariance of cosmic shear two-point functions is that from super-sample covariance~\citep[SSC;][]{2013PhRvD..87l3504T, 2018JCAP...10..053B, 2021A&A...646A.129J}. The SSC covariance can be approximated as a rank-1 update to the total covariance matrix~\citep{2019A&A...624A..61L}, with degradations to parameter constraints roughly determined by the alignment of the power spectrum response to a large-scale density fluctuation with the response to the parameter of interest. For the projected matter power spectrum,~\citet{2019A&A...624A..61L} find that $H_0$ has the smallest overlap with the SSC mode amongst the $\Lambda$CDM parameters, but a complete treatment of the SSC covariance for weak lensing is required to quantify the residual variance.} and neglect of systematics, and so the forecast constraints should be considered as lower limits.

\begin{figure*}
  \includegraphics[width=\textwidth]{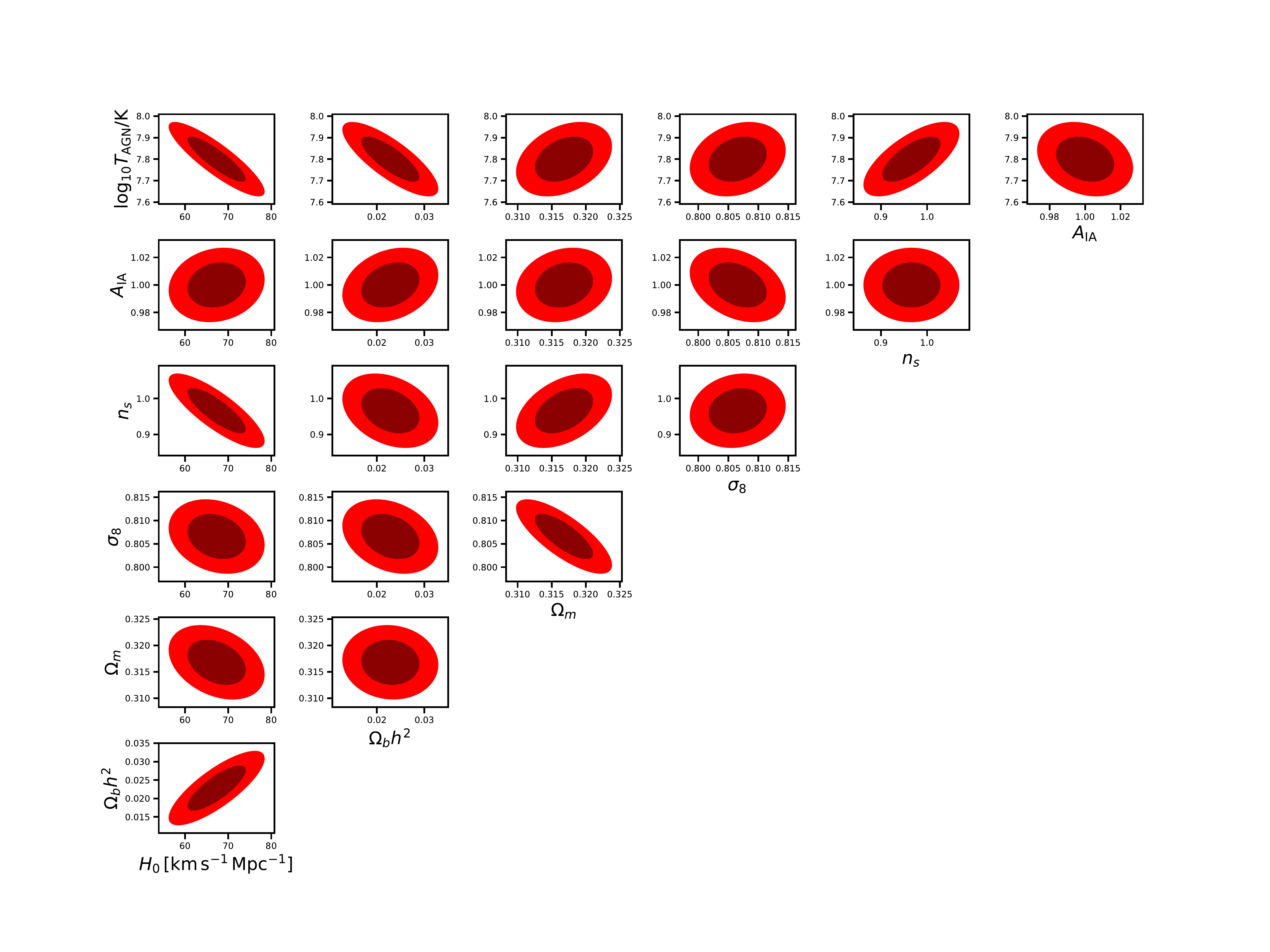}
  \caption{Forecast 1$\sigma$ and 2$\sigma$ constraints on $\Lambda$CDM parameters from a Fisher matrix forecast of a Euclid-like lensing survey. The parameter $\log_{10} T_{\mathrm{AGN}}/\mathrm{K}$ controls the amplitude of baryon feedback.}
  \label{fig:Euclid_2dellipses}
\end{figure*}

In Figure~\ref{fig:Euclid_2dellipses} we plot the 1$\sigma$ and 2$\sigma$ constraints on cosmological parameters expected from our toy Euclid-like survey. Several familiar degeneracy directions are apparent, such as the negative correlation of $\sigma_8$ and $\Omega_m$ and the positive corelation between the baryon feedback amplitude and $n_s$ due to their opposite effects on the small-scale matter power spectrum. Constraints on all parameters are generally very tight due to the high statistical constraining power of this toy survey, but intriguingly the forecast $H_0$ constraint is only $\sim 7\%$, i.e.~not competitive with even current measurements. The conditional error on $H_0$ fixing all other parameters is $0.16\%$, suggesting that degeneracies may severely limit the ability of future surveys to measure $H_0$. This gives strong support to the results of Section~\ref{sec:WLcosmodep} where we argued that $H_0$ information can only come from information in the power spectra beyond the power-law region $10^2 \lesssim l \lesssim 10^3$, requiring the breaking of degeneracies between $\Omega_m$ and $\Omega_m h^2$. This degeneracy breaking is inhibited by other parameters that control the shape of the matter spectrum such as $n_s$, $\Omega_b h^2$, and baryon feedback, and indeed these three parameters show the strongest degeneracies with $H_0$ in Figure~\ref{fig:Euclid_2dellipses}.

\begin{figure}
  \includegraphics[width=\columnwidth]{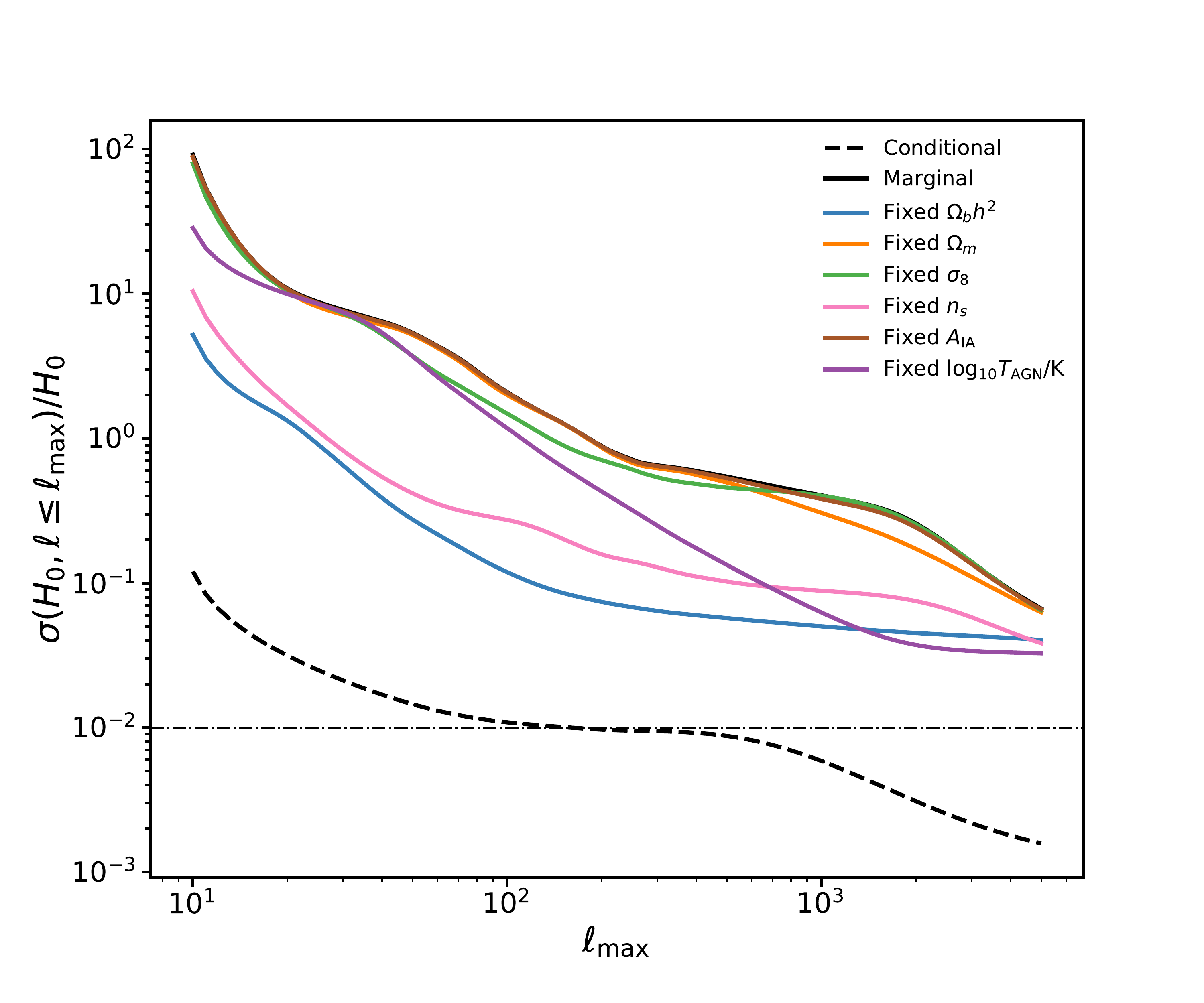}
  \caption{Forecast 1$\sigma$ fractional uncertainty on $H_0$ as a function of the maximum multipole included in our forecast, in the case that all other parameters are fixed (`conditional' error, black dashed curve), all other parameters are marginalised over (`marginal' error, black solid curve), or where each individual parameter in turn is held fixed (coloured solid curves). A baseline 1\% accuracy level is marked by the thin dot-dashed horizontal line.}
  \label{fig:Euclid_H0_fixing}
\end{figure}

The degeneracy structure of the Fisher matrix is also shown in Figure~\ref{fig:Euclid_H0_fixing} where we show the fractional error on $H_0$ as a function of $\ell_{\mathrm{max}}$ fixing each of the other parameters in turn. The conditional error on $H_0$ (fixing all other parameters in this space) is also shown, and exhibits a plateau between $10^2 \lesssim \ell \lesssim 10^3$ in accordance with the results of Section~\ref{sec:WLcosmodep}, although this is not particularly meaningful since it depends on the specific model parametrization. The ratio of the fully marginalised error on $H_0$ to its conditional error generally falls with $\ell_{\mathrm{max}}$ as degeneracies are broken, but stalls at $20 \lesssim l_{\mathrm{max}} \lesssim 60$ where the power spectrum is roughly a pure power law and then again at $300 \lesssim l_{\mathrm{max}} \lesssim 3000$. When only scales $\ell_{\mathrm{max}} \lesssim 700$ are used the limiting degeneracies are with $n_s$ and $\Omega_b h^2$ due to the similar effects of these parameters on the slope of the linear power spectrum, whereas at smaller scales baryon feedback is the limiting degeneracy due to its similar effect on the slope of the 1-halo term.

\begin{figure}
  \includegraphics[width=\columnwidth]{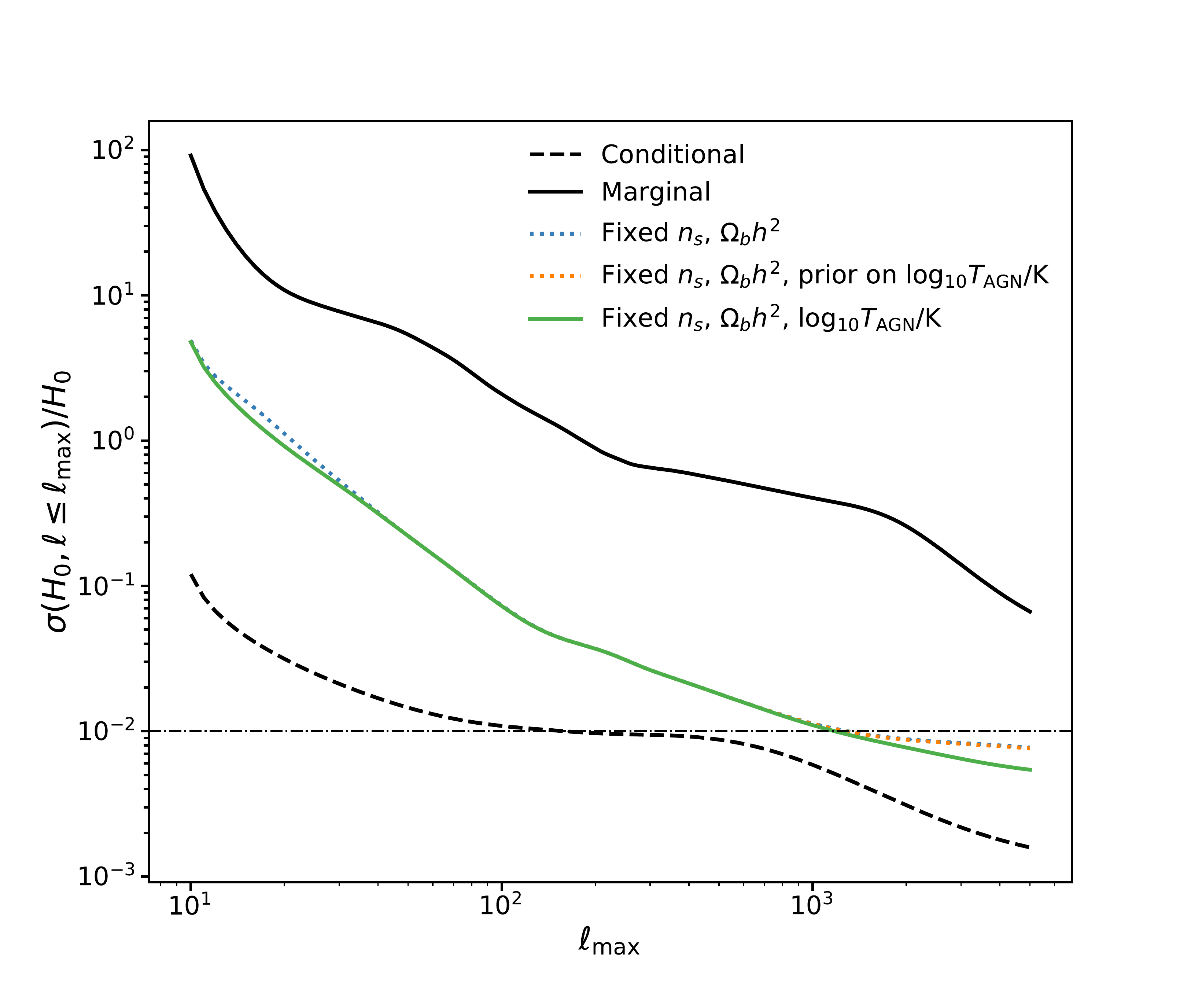}
  \caption{Same as Figure~\ref{fig:Euclid_H0_fixing} but fixing combinations of parameters that contribute most to a degeneracy with $H_0$. Note that imposing priors on $n_s$ and $\Omega_b h^2$ corresponding to the Planck 2018 marginalised constraints on each parameter individually give almost the same result as fixing them.}
  \label{fig:Euclid_H0_fix2pars}
\end{figure}

Figure~\ref{fig:Euclid_H0_fixing} shows that fixing any single parameter is not enough to get the marginalised error on $H_0$ below 1\%. In Figure~\ref{fig:Euclid_H0_fix2pars} we show how the forecast error on $H_0$ behaves with $\ell_{\mathrm{max}}$ when combinations of parameters are fixed. Sub-percent constraints are now possible if $\ell_{\mathrm{max}} \gtrsim 10^3$ if $n_s$ and $\Omega_b h^2$ are fixed, with additional gains if a prior on the baryon feedback parameter motivated from hydrodynamical simulations, as prescribed in~\citet{2021MNRAS.502.1401M}, is imposed\footnote{Note that the prior recommended in \citet{2021MNRAS.502.1401M} is uniform between $7.6 < \log_{10} T_{\mathrm{AGN}}/\mathrm{K} < 8.0$. To incorporate this into our Fisher matrix we replace this with a Gaussian having the same mean and variance.}. We find almost the same results if instead of fixing $n_s$ and $\Omega_b h^2$ we impose priors with the marginalised errors on each parameter reported by~\citet{2020A&A...641A...6P}.

We thus see that informative priors on both $n_s$ and $\Omega_b h^2$ will be required for sub-percent $H_0$ constraints with cosmic shear alone in our toy survey. This point is made further in Appendix~\ref{app:Euclid_with_prior}, where instead of fixing $n_s$, $\Omega_b h^2$ we impose the `lensing only' Planck priors used in Section~\ref{sec:lensing_only} and given in Table~\ref{tab:priors}. Figure~\ref{fig:Euclid_H0_priors} shows that these priors are not sufficiently informative to give sub-percent $H_0$ constraints. We find that a 1\% constraint is possible if $n_s$ is known to the current Planck precision and a BBN prior on $\Omega_b h^2$ is imposed (with or without a prior on baryon feedback), but only if all modes out to $\ell_{{\rm max}} = 5000$ are used, which we consider optimistic.

Figure~\ref{fig:Euclid_H0_priors} also shows constraints assuming $\ell_{{\rm max}} = 5000$ and lowering the minimum scale included in the analysis. Sub-percent constraints on $H_0$ are reached by $\ell_{{\rm min}} < 1000$, showing that it is not necessary to measure large scales to get $H_0$ with sub-percent in this toy survey -- it is sufficient to probe the 1-halo regime where enough modes are in principle available to break parameter degeneracies, although only if external information on $n_s$ and $\Omega_bh^2$ is provided. We caution however that off-diagonal terms in the covariance matrix will lower the constraining power across these scales. In particular we note that a measurement of $\ell_{\mathrm{eq}}$ is not required in this scenario.
%

We also investigated $H_0$ forecasts allowing the total neutrino mass to vary. We find that $H_0$ has a positive degeneracy with neutrino mass due to their opposite effects on small-scale lensing power (see Figure~\ref{fig:Cl_KV450_dependcies}). This tightens the requirements on any external information on other parameters that needs to be included to get $H_0$ to sub-percent, such that it is no longer sufficient to fix $n_s$ and $\Omega_b h^2$. Instead, we find that a tight prior on the baryon feedback amplitude is required for sub-percent $H_0$, and at least $\ell_{{\rm max}} > 2000$. This prior must be narrower (in terms of allowed range of values the parameter $\log_{10} T_{\mathrm{AGN}}/\mathrm{K}$ is allowed to take) than the simulation-based prior of~\citet{2021MNRAS.502.1401M} by at least a factor of three. This prior must be imposed additionally to informative priors on $n_s$ and $\Omega_b h^2$ with width comparable to their constraints from the CMB. Note that a BBN prior on $\Omega_b h^2$ is now not sufficient, even if tight $n_s$ and baryon feedback priors are imposed.

We thus conclude from this section that sub-percent constraints on $H_0$ with lensing alone will be very challenging even with the high statistical precision offered by forthcoming surveys. External priors on parameters controlling the slope of the power spectrum, such as $n_s$, $\Omega_b h^2$, and the baryon feedback amplitude will all be required, with the prior on $n_s$ in particular required to be highly informative. The lack of features in the lensing power spectrum is the primary hindrance to precision $H_0$ with lensing. A constraint is only possible due to the subtle effects of $\Omega_m h^2$ on the shape of the power spectrum - this needs to be distinguished from similar broad-band effects from other parameters if $H_0$ is to be measured well.

\section{Conclusions}
\label{sec:conc}

We have conducted a thorough study of the cosmological constraining power of weak lensing, paying special attention to the Hubble constant. We have studied the cosmological constraints that current galaxy and CMB lensing surveys can provide separately, in combination with each other, and in combination with BAO measurements. We have investigated the sensitivity of galaxy lensing two-point functions to various cosmological parameters in $\Lambda$CDM, within the framework of the halo model. Finally we looked at potential constraints from forthcoming lensing surveys on $\Lambda$CDM parameters, in particular $H_0$. The main findings of this work are as follows:

\begin{itemize}
\item
  Current lensing surveys alone do not provide useful constraints on $H_0$. The combination of galaxy and CMB lensing does however allow $\Omega_m$ to be constrained due to their different degeneracies with $\sigma_8$. Combining Planck lensing with KV450 we find $\Omega_m = 0.22 \pm 0.04$ and $\sigma_8 = 0.88 \pm 0.05$, with a residual degeneracy between these two parameters. Using instead SPTpol lensing with KV450 gives highly consistent results, $\Omega_m = 0.22 \pm 0.03$ and $\sigma_8 = 0.88 \pm 0.04$ assuming a narrow prior on $n_s$. These results are consistent with recent measurements from KiDS-1000~\citep{2020arXiv201016416T}.
\item
  Constraints in the $\Omega_m$-$H_0$ plane are dominated by Planck lensing due its measurement of the equality scale, so combining Planck lensing with KV450 and BAO+BBN gives an $H_0$ constraint comparable to that from just Planck lensing and BAO+BBN. Out tightest constraint on $H_0$ comes from combining low and high redshift BAO+BBN, Planck lensing, and galaxy lensing, and is $H_0 = 67.4 \pm 0.9 \; \mathrm{km} \, \mathrm{s}^{-1} \, \mathrm{Mpc}^{-1}$ with a narrow prior on $n_s$, and $H_0 = 67.6 \pm 1.1 \; \mathrm{km} \, \mathrm{s}^{-1} \, \mathrm{Mpc}^{-1}$ with a broad prior on $n_s$. These constraints are independent of the primary CMB fluctuations and are $4.0\sigma$ and $3.6\sigma$ lower than the SH0ES measurement of~\citet{2019ApJ...876...85R} respectively. Constraining power is lost when the BBN prior on $\Omega_bh^2$ is dropped.
\item
  We have shown that current cosmic shear measurements are very insensitive to $H_0$. Using analytic arguments we were able to derive accurate parameter dependences of the 1-halo and 2-halo matter power spectrum, and hence the parameter dependence of the lensing correlation functions. We updated the results of~\citet{1997ApJ...484..560J} by showing how the dependence of lensing on the parameter combination $\sigma_8 \Omega_m^{0.5}$ follows from the halo model.
\item
  We showed that the scales and redshifts probes by current surveys are such that the shear power spectrum can be approximated by a power law with amplitude proportional to $S_8$. There is almost no sensitivity to $H_0$ at fixed $\sigma_8$ and $\Omega_m$ across a broad range of scales due to the length scales of the halo model being tied to the horizon scale, and the fact that $\sigma_8$ is roughly unity. $H_0$ dependence can emerge from measurements of large linear scales around the peak of the lensing power spectrum, or from measurements of small scales deep in the 1-halo regime. These features taken together explain why current surveys are so insensitive to $H_0$. Angular scales such as the linear-nonlinear transition scale or the angular size of the virial radius of a typical NFW halo do not impart strong dependence on $H_0$ due to their lack of sharpness and sensitivity to baryon feedback. $H_0$ affects small and large scales in an opposite way, and as such is partially degenerate with other parameters that change the slope of the power spectrum, such as $n_s$ and $\Omega_b h^2$.
\item
  We examined the potential of forthcoming lensing surveys to constrain $H_0$. A toy Euclid-like survey only constrains $H_0$ with $7\%$ precision after marginalising over other parameters. The main degeneracies are parameters that change the slope of the power spectrum, specifically $n_s$, $\Omega_b h^2$, and the baryon feedback amplitude. We showed that a tight prior on $n_s$, comparable with current constraints from the CMB, is necessary to measure $H_0$ with sub-percent precision, as well as a prior on $\Omega_b h^2$ at least as informative as that from BBN. Demands on the priors increase when neutrino mass is additionally allowed to vary, necessitating a tightening of baryon feedback prior by a factor of three over current bounds from hydrodynamic N-body simulations.
\end{itemize}

Our focus in this work has been on parameter constraints from the two-point function of lensing maps. The shear signal is non-Gaussian, so there is considerable information contained in the higher-order cumulants and other descriptors preserving more of the information content. Folding in information from, say, the lensing bispectrum or convergence peak counts may well break further degeneracies and improve constraints on $H_0$. This is an interesting avenue for future study.

One of the main motivations for this work was the observation that CMB lensing combined with galaxy lensing can give a constraint on $\Omega_m$, which is all that is needed for a CMB-independent constrain on $H_0$ when combined with BAO+BBN. It turned out this combination was no more constraining than CMB lensing alone combined with BAO+BBN, due to the measurement of $L_{\mathrm{eq}} \propto \Omega_m^{0.6}h$ in Planck lensing. The aim for future galaxy lensing surveys should be a precise measurement of this scale, which should provide a constraint on $\Omega_m h$, and hence a constraint on $H_0$ from lensing \emph{alone}. This would be a constraint independent of either CMB or BAO. If the $H_0$ tension persists, such a measurement could be very valuable for disentangling systematics from new physics.
%

\section*{Acknowledgements}

I thank Federico Bianchini for sharing the SPTpol lensing chains and Alexander Mead for useful conversations and for sharing his halo model code. I thank Catherine Heymans, Weikang Lin, Alexander Mead, John Peacock, Adam Riess, and Andy Taylor for comments on the manuscript. Some of the results of this paper are based on data products from observations made with ESO Telescopes at the La Silla Paranal Observatory under programme IDs 177.A-3016, 177.A-3017, 177.A-3018, 179.A-2004, and 298.A-5015.

\section*{Data Availability}

The data used in this article are all publicly available or available upon request from the relevant parties. Planck chains and likelihoods are available from the ~\href{https://pla.esac.esa.int/#home}{ESA Planck Legacy Archive}. Data products from the Kilo-Degree Survey are available from~\url{http://kids.strw.leidenuniv.nl/}. SPTpol data products are available from~\url{https://pole.uchicago.edu/public/data/lensing19/}. All other data sets and likelihoods are available from the~\href{https://github.com/brinckmann/montepython_public}{\textsc{MontePython} GitHub repository}.



\bibliographystyle{mnras}
\bibliography{references} 




\vspace{-0.5cm}

\appendix

\section{Choosing different priors in KV450}
\label{app:KV450_vary_priors}

\begin{figure}
  \includegraphics[width=\columnwidth]{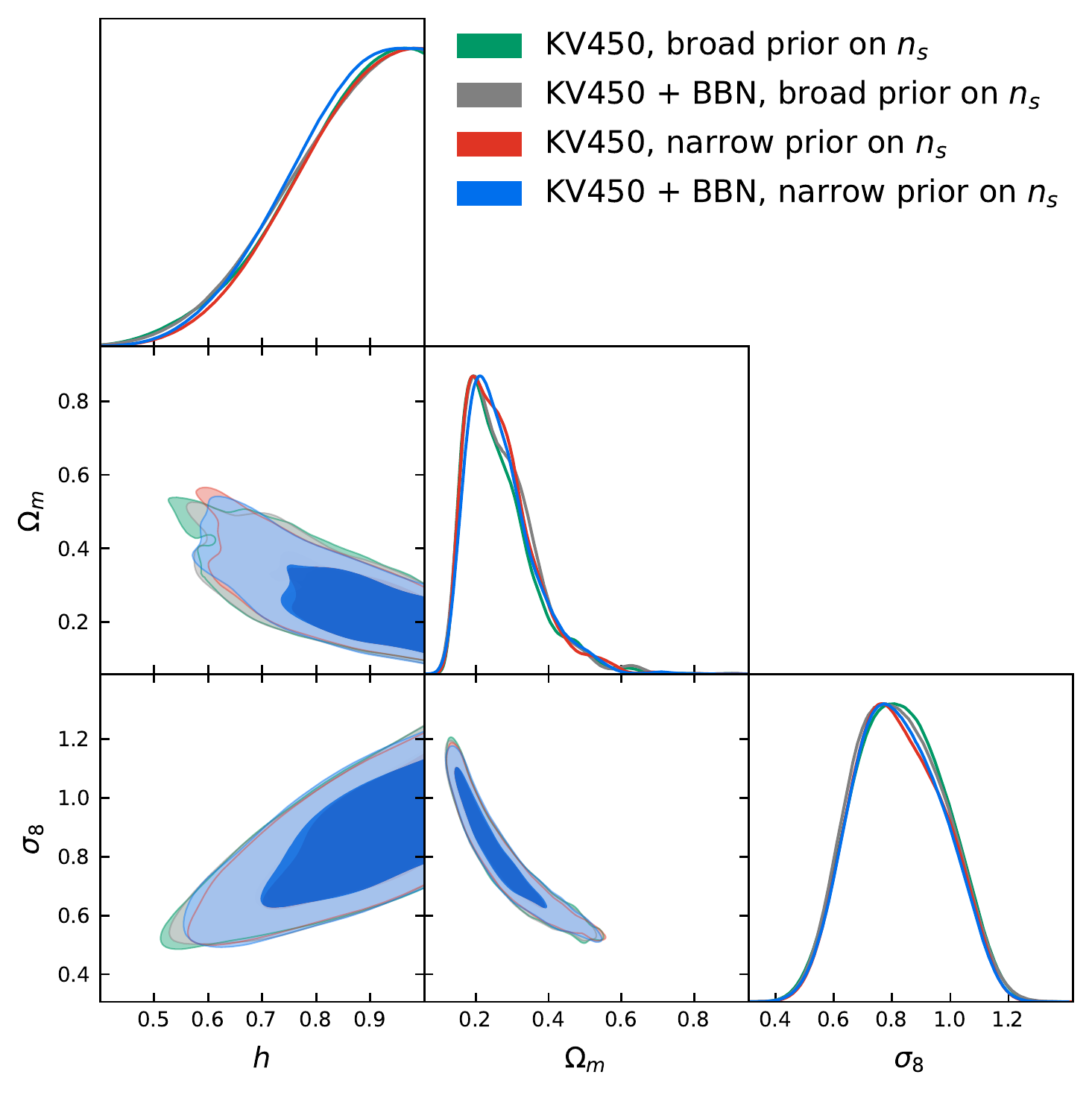}
  \caption{68\% and 95\% posterior credible regions on the parameters $\Omega_m$, $\sigma_8$, and $h$ from the weak lensing correlation functions measured in KV450, with different choices of prior. `+BBN' in this plot refers to a BBN prior on $\Omega_b h^2$.}
  \label{fig:KV450_3params}
\end{figure}

In Figure~\ref{fig:KV450_3params} we show posterior constraints on the parameters $\Omega_m$, $\sigma_8$, and $h$ inferred from KV450, assuming various combinations of prior on $n_s$ or $\Omega_b h^2$. For $n_s$ the prior is either a Gaussian with $n_s = 0.96 \pm 0.02$ as in the `Planck lensing' priors given in Table~\ref{tab:priors}, or uniform in the range $[0.7,1.3]$ as in the `KV450' priors given in Table~\ref{tab:priors}. These are referred to as `narrow' and `broad' respectively in Figure~\ref{fig:KV450_3params}. The prior on $\Omega_bh^2$ is either a BBN prior given by a Gaussian with $\Omega_b h^2 = 0.0222 \pm 0.0005$, or the original prior used in~\citet{2020A&A...633A..69H} of a uniform prior in the range $[0.019, 0.026]$.

Figure~\ref{fig:KV450_3params} demonstrates that the posterior constraints from KV450 in this parameter space are insensitive to these various choices of prior. The small residual differences are likely comparable to the error in the nested sampling contours from the finite number of live points used, and in any case are negligible compared with the width of the contours.
%

\section{Weak lensing + BBN as a standard ruler calibrator}
\label{sec:lensing_bao}

As anticipated, the combination of CMB lensing with galaxy lensing does not improve constraints on $H_0$ significantly due to the similar degeneracy directions in the individual posteriors. Additional prior information is required to measure $H_0$. For example,~\citet{2021MNRAS.501.1823B} combine Planck CMB lensing, an external constraint on $\Omega_m$ from the Pantheon survey of Type 1a supernovae, and a prior on the initial power spectrum amplitude $A_s$, breaking the degeneracies shown in Figure~\ref{fig:KV450_Pns_BBN_triangle} to obtain $h = 0.735 \pm 0.053$, a 7\% measurement consistent with both primary CMB and local distance ladder results. However, as shown in~\citet{2020A&A...641A...8P} and suggested by Figure~\ref{fig:KV450_Pns_BBN_triangle}, lensing provides a constraint in the $\Omega_m$-$h$ plane which may be used to calibrate the BAO scale. This provides a constraint on $H_0$ independent from the primary CMB fluctuations, in the spirit of the measurements of~\citet{2013MNRAS.436.1674A, 2015PhRvD..92l3516A}.

BAO experiments with high enough signal-to-noise are able to measure both the transverse BAO scale, $\theta_d(z_i) \equiv \chi(z_i)/r_d$, and the radial BAO scale $\delta z_d \equiv H(z_i)^{-1}/r_d$ at a range of redshifts $z_i$, where $r_d$ is the comoving sound horizon at the drag epoch given in $\Lambda$CDM by
\begin{equation}
  r_d = \int_{z_d}^\infty \frac{\mathrm{d}z}{H(z)} \left[3 + \frac{9}{4}\frac{\rho_b(z)}{\rho_\gamma(z)}\right]^{-1/2},
  \label{eq:rd}
\end{equation}
where $z_d$ is the drag epoch redshift, $\rho_{\gamma}$ is the background CMB energy density (fixed by the CMB temperature), $\rho_b$ is the baryon energy density (proportional to $\Omega_b h^2$) and $H(z)$ is the Hubble parameter. Since $z_d$ depends only weakly on parameters, $r_d$ is mostly a function of the high-redshift $H(z)$ and $\Omega_bh^2$. The former is mostly sensitive to $\Omega_m h^2$, so to high accuracy in $\Lambda$CDM we have $r_d = r_d\left(\Omega_m h^2, \Omega_b h^2\right)$. Since $\Lambda$ is dynamically important at and below the redshifts where BAO are actually measured both $\chi$ and $H$ are functions of $\Omega_m$ and $h$ separately, so both the transverse and radial BAO scales are functions of $\Omega_m$, $h$, and $\Omega_b h^2$ separately. The baryon energy density is tightly constrained by our BBN prior, so the BAO measurements considered here constrain partially degenerate combinations of $\Omega_m$ and $h$.

\begin{figure*}
  \includegraphics[width=\columnwidth]{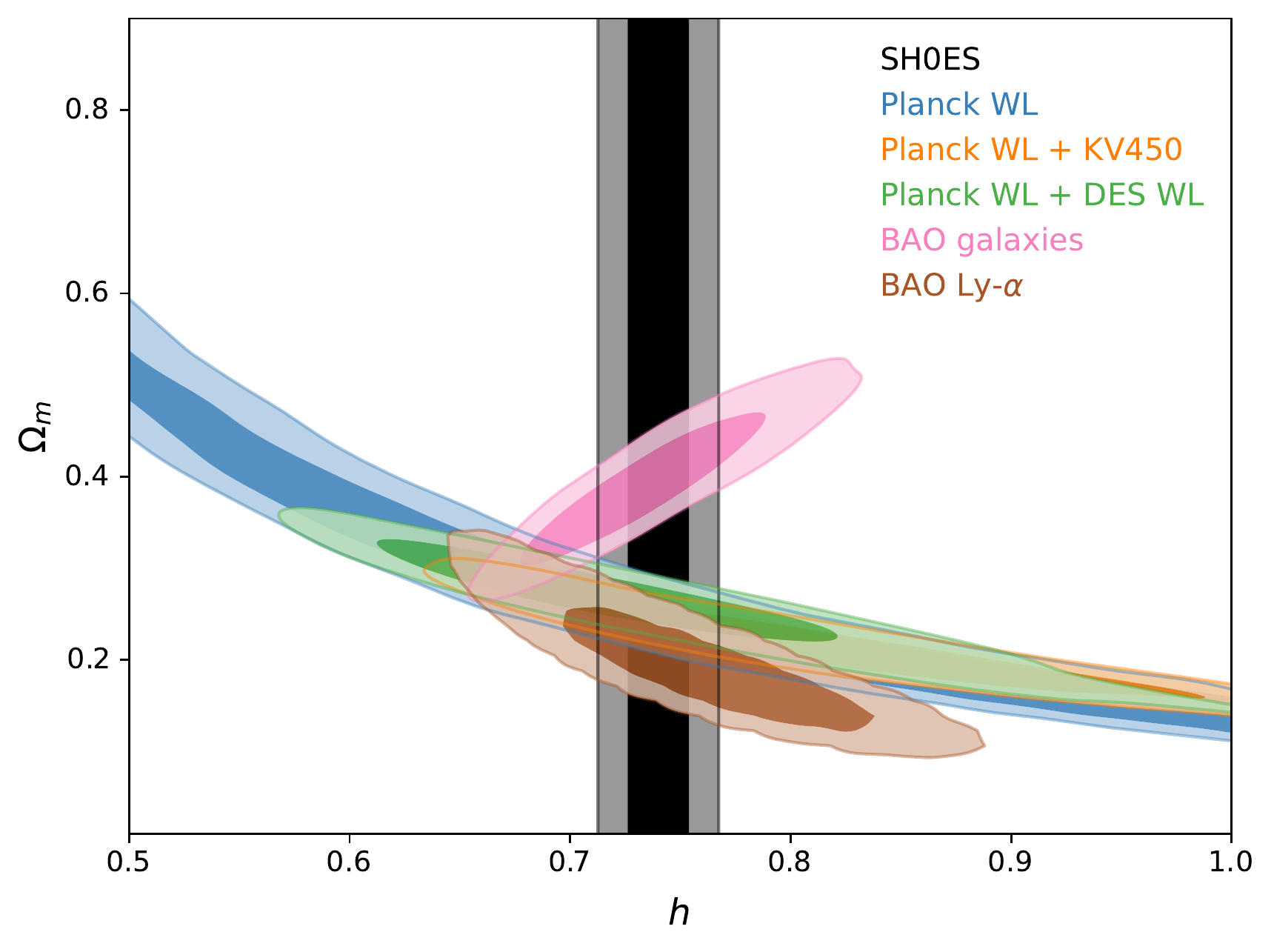}
  \includegraphics[width=\columnwidth]{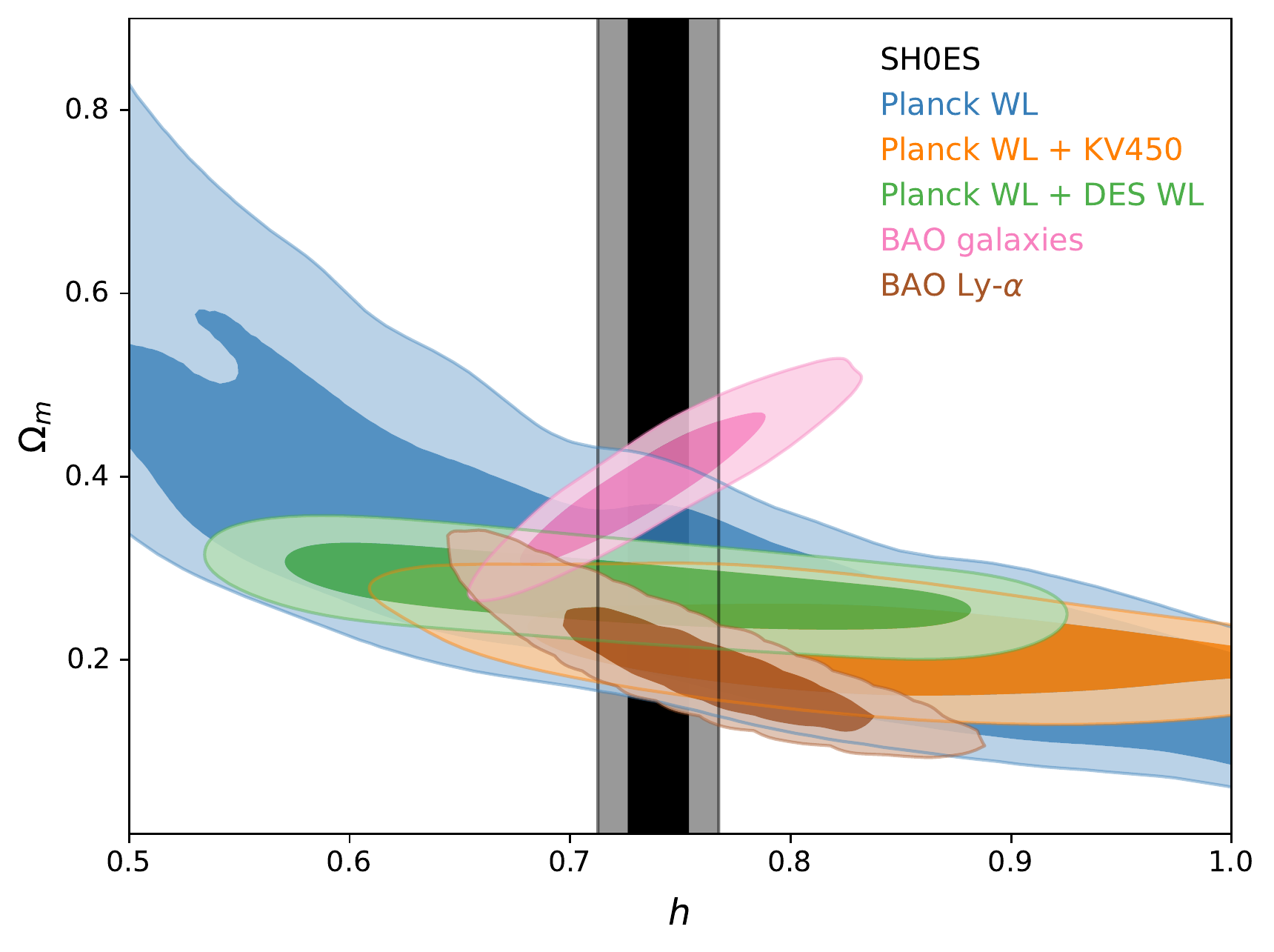}
  \caption{Marginalised 68\% and 95\% constraints on $\Omega_m$ and $h$ from various lensing and BAO data assuming a narrow (left panel) or broad (right panel) prior on $n_s$. Adding galaxy weak lensing to Planck lensing tightens constraints on $\Omega_m$ considerably (see Figure~\ref{fig:bananas}), and additionally help to rule out low values of $h$. The combination of Planck lensing with galaxy lensing (orange and green) gives constraints in this parameter space consistent with galaxy BAO (pink) and Ly-$\alpha$ BAO (brown), particularly true when a narrow $n_s$ prior is imposed. The SH0ES constraint is shown in black. A BBN prior on $\Omega_b h^2$ has been imposed for each analysis except in the left panel for the combination including DES, which uses the `DES priors' described in the text.}
  \label{fig:Omegam_h_changing_ns_prior}
\end{figure*}

In Figure~\ref{fig:Omegam_h_changing_ns_prior} we show constraints in the $\Omega_m$-$h$ plane from a compilation of BAO measurements using galaxy redshifts (labelled `BAO galaxies') and measurements of the BAO scale using the Lyman-$\alpha$ forest auto spectrum and its cross-correlation with quasars (labelled `BAO Ly-$\alpha$'). We also show the distance ladder constraint from SH0ES~\citep{2019ApJ...876...85R} in black, corresponding to $H_0 = 74.0 \pm 1.4 \; \mathrm{km} \, {\rm s}^{-1} \, {\rm Mpc}^{-1}$. For our galaxy BAO measurements we use the anisotropic measurements from BOSS DR12~\citep{2017MNRAS.470.2617A} combined with low-redshift measurements of the spherically-averaged BAO scale from the 6dF Galaxy Survey~\citep{2011MNRAS.416.3017B} and the SDSS DR7 Main Galaxy Sample~\citep{2015MNRAS.449..835R}. For our Lyman-$\alpha$ forest BAO measurements we use the likelihood of~\citet{2019JCAP...10..044C} which uses measurements of the BAO scale in the autospectrum of Lyman-$\alpha$ absorption features in eBOSS DR14~\citep{2019A&A...629A..85D} and from the cross-correlation of Lyman-$\alpha$ absorption with quasars in eBOSS DR14~\citep{2019A&A...629A..86B}, both of which measure BAO at high redshift ($z \approx 2.35$)\footnote{Recently, eBOSS DR16~\citep{2020arXiv200708991E} released galaxy and quasar BAO measurements in the range $0.6 \lesssim z \lesssim 2.2$, filling the redshift gap between BOSS DR12 ($0.2 \lesssim z \lesssim 0.6$) and the high redshift eBOSS Lyman-$\alpha$ measurements ($z \approx 2.35$). Although we do not include these more recent measurements in our analysis, we do not expect our constraints to change significantly given the modest increase in constraining power in the $\Omega_m$-$h$ plane from the combined low and high redshift BAO measurements (1.8\% pre-DR16 to 1.4\% post-DR16 on $H_0$ with BAO+BBN alone). The main effect of including the DR16 data is to rule out $\Omega_m \lesssim 0.2$ at 95\% confidence in the high-redshift ($z>1$) sample, and to increase the consistency of $z<1$ and $z>1$ constraints for $\Lambda$CDM models. As the main focus of this paper is a study of how $H_0$ may be constrained with lensing, we believe our omission of the eBOSS DR16 data is justified.}.

Constraints from galaxy weak lensing alone in this parameter space are too broad to give useful information on $H_0$\footnote{As shown in~\citet{2020arXiv201004158J}, galaxy lensing can help to constrain models with high $\Omega_m h^2$ (and hence low sound horizon) if one is willing to combine Planck primary CMB measurements and cosmic shear in a joint analysis. Such models are favoured by trying to simultaneously fit the SH0ES measurement of $H_0$ and the angular size of the sound horizon measured with Planck and BAO. \citet{2020arXiv201004158J} show that the $S_8$ implied by Planck in such models is several sigma away from that measured by lensing (a similar result was found in the context of Early Dark Energy models in ~\citealt{PhysRevD.102.043507}). This can be understood by noting that $S_8 \propto A_s^{0.5} \omega_m^{1.25} h^{-0.75}$ at fixed $n_s$ and $\Omega_b h^2$, so roughly speaking fixed $A_s$ implies a higher $S_8$ when $\omega_m$ is higher. Planck's $S_8$ is already slightly high compared with that of galaxy lensing and these models exacerbate the tension. Note that in this work we do not combine with any primary CMB measurements, and use only data at low redshift.}, but the combination of BAO with CMB lensing is enough to give tight constraints on $H_0$ when a BBN prior on $\Omega_b h^2$ is imposed, as previously found in~\citet{2016A&A...594A..15P, 2020A&A...641A...8P, 2020ApJ...888..119B, 2020ApJ...904L..17P}. Figure~\ref{fig:Omegam_h_changing_ns_prior} demonstrates a nice consistency between the two sets of BAO constraints and Planck lensing, all three contours intersecting around similar values of $\Omega_m$ and $h$. The almost orthogonal degeneracy directions of the two BAO constraints are a result of the different redshifts being probed, and is discussed in detail in~\citet{2015PhRvD..92l3516A, 2018ApJ...853..119A, 2019JCAP...10..044C, 2019JCAP...10..029S}. Their combination implies values of $\Omega_m$ and $h$ giving a CMB lensing power spectrum peak aligned with that measured by Planck. The addition of galaxy weak lensing adds little to the CMB lensing once BAO are included, which could have been anticipated from the broad contours in this parameter space from galaxy lensing. When either KV450 or DES lensing~\citep{2018PhRvD..98d3526A} are combined with Planck lensing\footnote{For our DES+Planck constraints we use the public MCMC chains provided by the Planck collaboration.} models with very low $h \lesssim 0.6$ are excluded, as also seen in Figure~\ref{fig:KV450_Pns_BBN_triangle}.

In the right panel of Figure~\ref{fig:Omegam_h_changing_ns_prior} we show how the constraints loosen when the informative prior on $n_s$ is dropped. The Planck lensing constraints become substantially weaker in this parameter space due to new parameter degeneracies which leave the shape of the CMB lensing power spectrum fixed. The inclusion of galaxy weak lensing information is hence relatively more important, with the addition of DES lensing (green contours) bringing the weak lensing constraint back into close agreement with the combined BAO constraint. Instead using KV450 (orange contours) results in slightly less perfect overlap, but the combination is still less than 2$\sigma$ away from the combined BAO constraint.

\begin{figure}
  \includegraphics[width=\columnwidth]{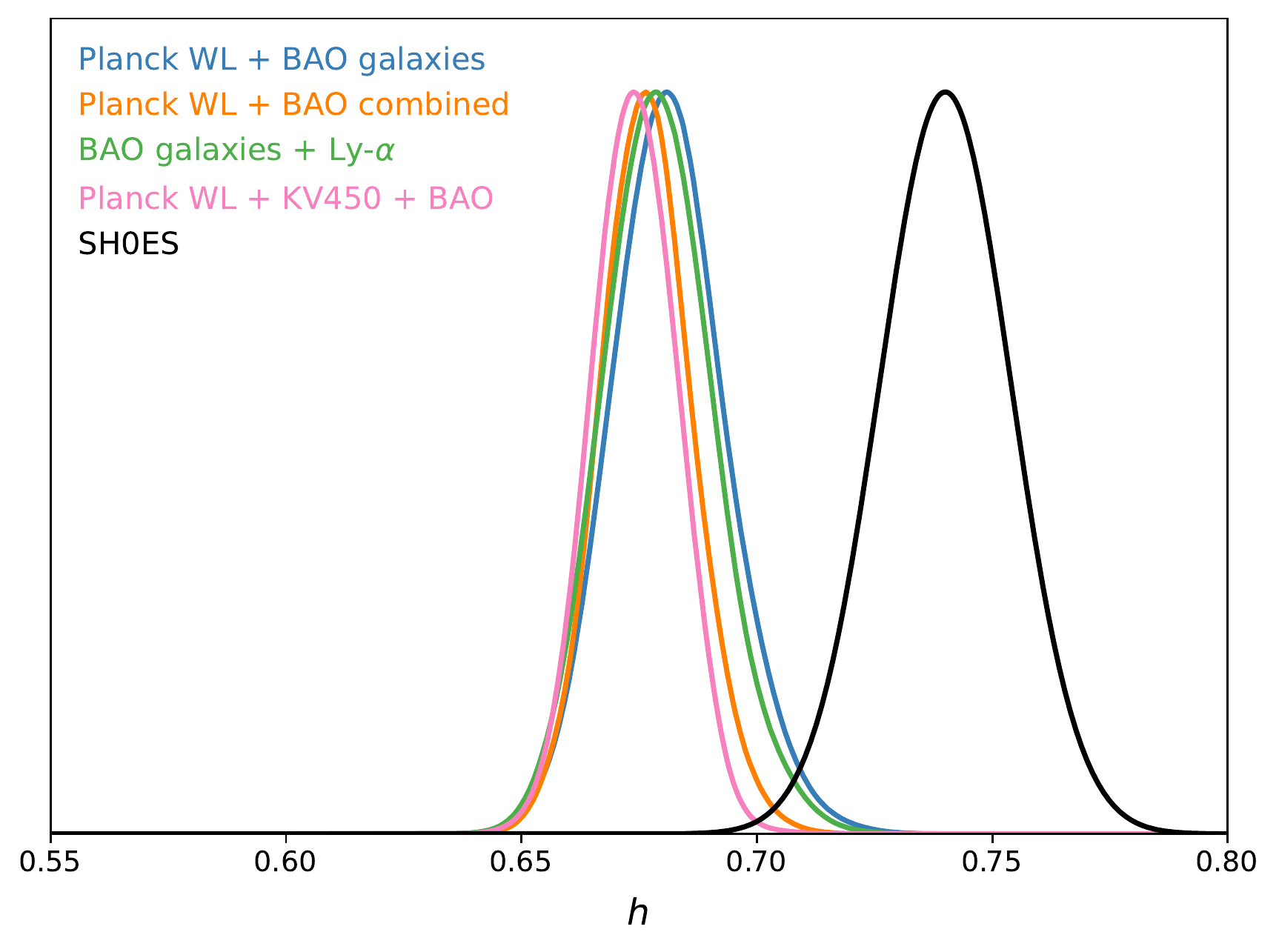}
  \includegraphics[width=\columnwidth]{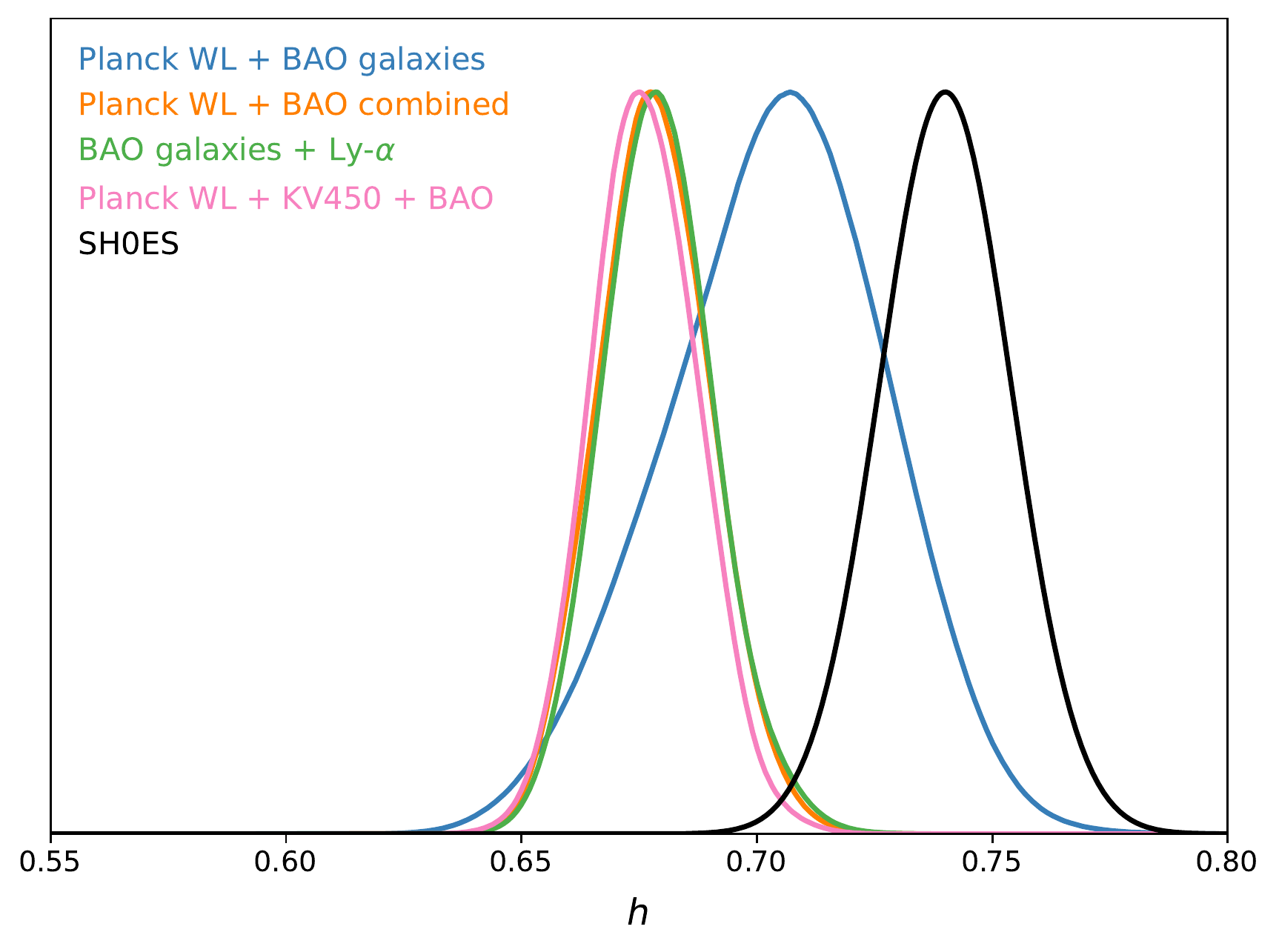}
  \caption{Constraints on $h$ from various lensing and BAO data assuming a narrow (top panel) or broad (bottom panel) prior on $n_s$. The reported Planck lensing + galaxy BAO constraint from~\citet{2020A&A...641A...8P} weakens significantly when the $n_s$ prior is relaxed, but this lost information is recovered when adding the Ly-$\alpha$ BAO (with the addition of galaxy weak lensing introducing a slight shift and tightening of the posterior). A BBN prior on $\Omega_b h^2$ has been imposed for each analysis.}
  \label{fig:H0_changing_ns_prior}
\end{figure}

The level of consistency between these data sets motivates constraining $H_0$ using their joint likelihood. In Figure~\ref{fig:H0_changing_ns_prior} we show the resulting constraints on $h$ from such combinations, with an informative (top panel) or uninformative (bottom panel) prior on $n_s$. The degeneracy direction in the $\Omega_m$-$h$ plane from Lyman-$\alpha$ BAO happens to be well aligned with that of CMB lensing (see Figure~\ref{fig:Omegam_h_changing_ns_prior}) so the most constraining pair of data sets is Planck lensing with galaxy BAO. With an informative prior on $n_s$, the combined constraints are (all constraints are plus BBN
\begin{align}
  H_0 &= 68.1^{+1.1}_{-1.3} \, {\rm km} \, {\rm s}^{-1} \, {\rm Mpc}^{-1} \, \text{(Planck lensing + BAO galaxies)}, \\
  H_0 &= 67.9^{+1.2}_{-1.2} \, {\rm km} \, {\rm s}^{-1} \, {\rm Mpc}^{-1} \, \text{(BAO galaxies + BAO Ly}\alpha \text{)}, \\
  H_0 &= 67.7^{+0.9}_{-1.0} \, {\rm km} \, {\rm s}^{-1} \, {\rm Mpc}^{-1} \, \text{(Planck lensing + all BAO)}, \\
  H_0 &= 67.4^{+0.9}_{-0.9} \, {\rm km} \, {\rm s}^{-1} \, {\rm Mpc}^{-1} \, \text{(Planck lensing + KV450 + all BAO)}.\label{eq:H0_all_Pns}
\end{align}
When the informative prior on $n_s$ is lifted, the lensing constraints weaken to
\begin{align}
  H_0 &= 70.4^{+2.5}_{-2.1} \, {\rm km} \, {\rm s}^{-1} \, {\rm Mpc}^{-1} \, \text{(Planck lensing + BAO galaxies)}, \\
  H_0 &= 67.8^{+1.2}_{-1.2} \, {\rm km} \, {\rm s}^{-1} \, {\rm Mpc}^{-1} \, \text{(Planck lensing + all BAO)}, \\
  H_0 &= 67.6^{+1.1}_{-1.1} \, {\rm km} \, {\rm s}^{-1} \, {\rm Mpc}^{-1} \, \text{(Planck lensing + KV450 + all BAO)}.\label{eq:H0_all_Kns}
\end{align}

As anticipated from Figure~\ref{fig:Omegam_h_changing_ns_prior}, the loss of information from lifting the prior on $n_s$ allows is mostly alleviated when adding both sets of BAO constraints. Using only the BAO measurements from galaxies, constraints on $H_0$ weaken by roughly a factor of two\footnote{When adopting DES priors instead of KV450 priors (see Table~\ref{tab:priors}), \citet{2020A&A...641A...8P} found that the $H_0$ constraint from Planck lensing plus galaxy BAO weakens to $68.0 \pm 1.5 \, {\rm km} \, {\rm s}^{-1} \, {\rm Mpc}^{-1}$ (see their Table 2), i.e. a less severe loss of information. The KV450 prior on $n_s$ is wider by a factor of three, suggesting even the broad DES prior on $n_s$ is adding significant information for the inference of $H_0$.}. This is comparable to the loss of information from allowing the neutrino mass $\sum m_\nu$ to vary, which gives $H_0 = 70.6^{+1.8}_{-2.4} \, {\rm km} \, {\rm s}^{-1} \, {\rm Mpc}^{-1}$ for Planck lensing + galaxy BAO with Planck priors. Massive neutrinos change the shape of the lensing power spectrum by suppressing the potential below their free streaming scale after they become non-relativistic~\citep{2006PhR...429..307L, 2012MNRAS.425.1170H}. This is partially degenerate with a change in $n_s$, and opens up more freedom in $\Omega_m$, $h$ and $\sigma_8$ to change the shape of the power spectrum at fixed amplitude. In the case of galaxy lensing, neutrino mass has non-trivial degeneracies with the baryon feedback model as well $\Lambda$CDM parameters that change the small-scale amplitude~\citep{2016MNRAS.459.1468M, 2020MNRAS.493.1640C}.

Our strongest constraint on $H_0$ with Planck lensing comes from the combination with BAO+BBN and KV450 (although the information is dominated by BAO and Planck lensing), given in Equations~\eqref{eq:H0_all_Pns} and~\eqref{eq:H0_all_Kns}. These constraints are $4.0\sigma$ and $3.6\sigma$ lower than the local measurement from~\citet{2019ApJ...876...85R} adopting an informative or uninformative $n_s$ prior respectively, and do not make use of primary CMB data at all except the temperature monopole. Moreover, these are consistent with primary CMB measurements of $H_0$ in $\Lambda$CDM from Planck, which give $H_0 = 67.27 \pm 0.60 \, {\rm km} \, {\rm s}^{-1} \, {\rm Mpc}^{-1}$ (TT,TE,EE+lowE measurements from ~\citealt{2020A&A...641A...6P}), $4.4\sigma$ lower than the local measurement. The BBN prior plays an important role here - dropping it yields $H_0 = 70.0^{+8.4}_{-4.6} \; \mathrm{km} \, \mathrm{s}^{-1} \, \mathrm{Mpc}^{-1}$, consistent with both Planck and SH0ES.

\begin{figure}
  \includegraphics[width=\columnwidth]{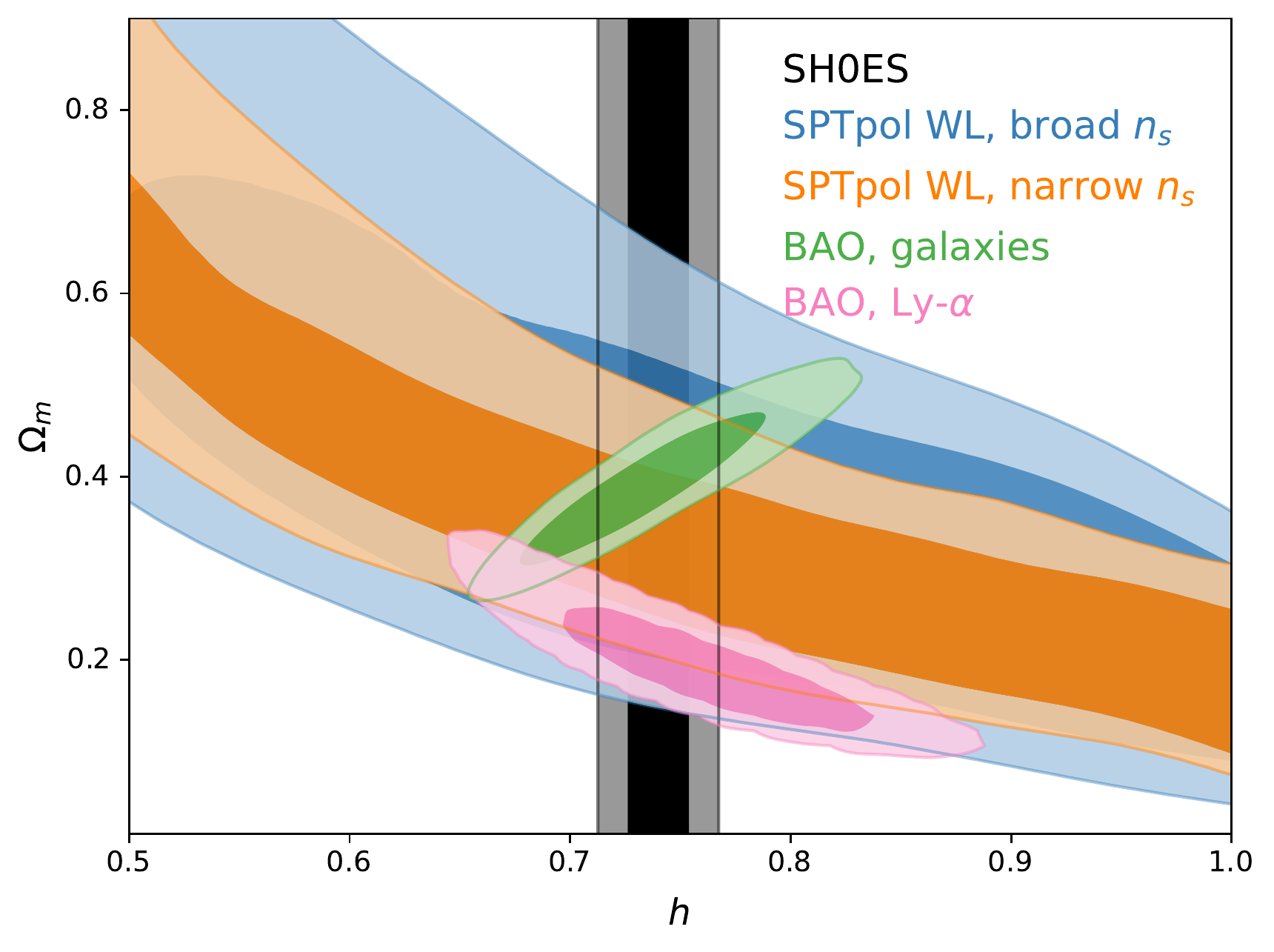}
  \caption{Marginalised 68\% and 95\% constraints on $\Omega_m$ and $h$ from BAO (green and pink) and CMB lensing with a broad or narrow prior on $n_s$ imposed (blue and orange respectively) using SPTpol for the CMB experiment. The SH0ES constraint is shown in black. Planck lensing exhibits excellent consistency with both galaxy BAO and Ly-$\alpha$ BAO in this parameter space. A BBN prior on $\Omega_b h^2$ has been imposed for each analysis.}
  \label{fig:Omegam_h_combined_changing_CMBWL}
\end{figure}

In Figure~\ref{fig:Omegam_h_combined_changing_CMBWL} we show constraints in the $\Omega_m$-$h$ plane swapping Planck's lensing power spectrum for that of SPTpol. As discussed above, the width of the contours in the well constrained direction reflects the accuracy with which the peak in the deflection angle power spectrum has been measured, so constraints from combining SPTpol with BAO+BBN are not as powerful at constraining $H_0$ compared with Planck. \citet{2020ApJ...888..119B} found that combining galaxy BAO+BBN with SPTpol lensing and an informative $n_s$ prior gives $H_0 = 72.0^{+2.1}_{-2.5} \, {\rm km} \, {\rm s}^{-1} \, {\rm Mpc}^{-1}$, i.e. almost double the uncertainty compared to Planck lensing plus BAO. This constraint relaxes further when we drop the informative prior on $n_s$ to $H_0 = 73.1 \pm 3.0  \, {\rm km} \, {\rm s}^{-1} \, {\rm Mpc}^{-1}$. As described in \citet{2020ApJ...888..119B} and evident from Figures~\ref{fig:Omegam_h_changing_ns_prior} and~\ref{fig:Omegam_h_combined_changing_CMBWL}, the SPTpol lensing-only constraints intersect the galaxy BAO contours at higher values of $H_0$ than Planck, although the two are still statistically consistent. This is consistent with the low lensing power in SPTpol compared with the best-fitting Planck model on the largest scales (see Figure 4 of~\citealt{2019ApJ...884...70W}) which shifts the inferred peak in the deflection angle power spectrum to slightly smaller angular scales, i.e. a higher value of $\Omega_m^{0.6}h$. This also improves the overlap between the parameter contours from SPTpol lensing-only and BOSS DR12 BAO compared with Planck lensing-only, driven on the galaxy side by the preference of line-of-sight BAO for higher $H_0$~\citep{2020PhRvD.102b3510W}.

To summarise this section, we have confirmed previous results that current weak lensing data (from the CMB and galaxies) on its own is unable to place constraints on $H_0$, due to a degeneracy with $\Omega_m$. Combining with BAO and a prior on $\Omega_b h^2$ breaks this degeneracy, and we have extended previous results by showing that the resultant constraint on $H_0$ is sensitive to the prior on $n_s$ that is imposed, with error bars inflating by a factor of two unless BAO at widely separated redshifts are included. Cosmological constraints from low-redshift BAO, high-redshift BAO, and CMB lensing (either from Planck or SPTpol) are consistent, leading to an improved constraint on $H_0$ that is between $3.6\sigma$ and $4\sigma$ lower than the local measurement of~\citet{2019ApJ...876...85R} when a BBN prior is imposed.

\section{Euclid-like constraints on $H_0$ with priors}
\label{app:Euclid_with_prior}

\begin{figure*}
  \includegraphics[width=\textwidth]{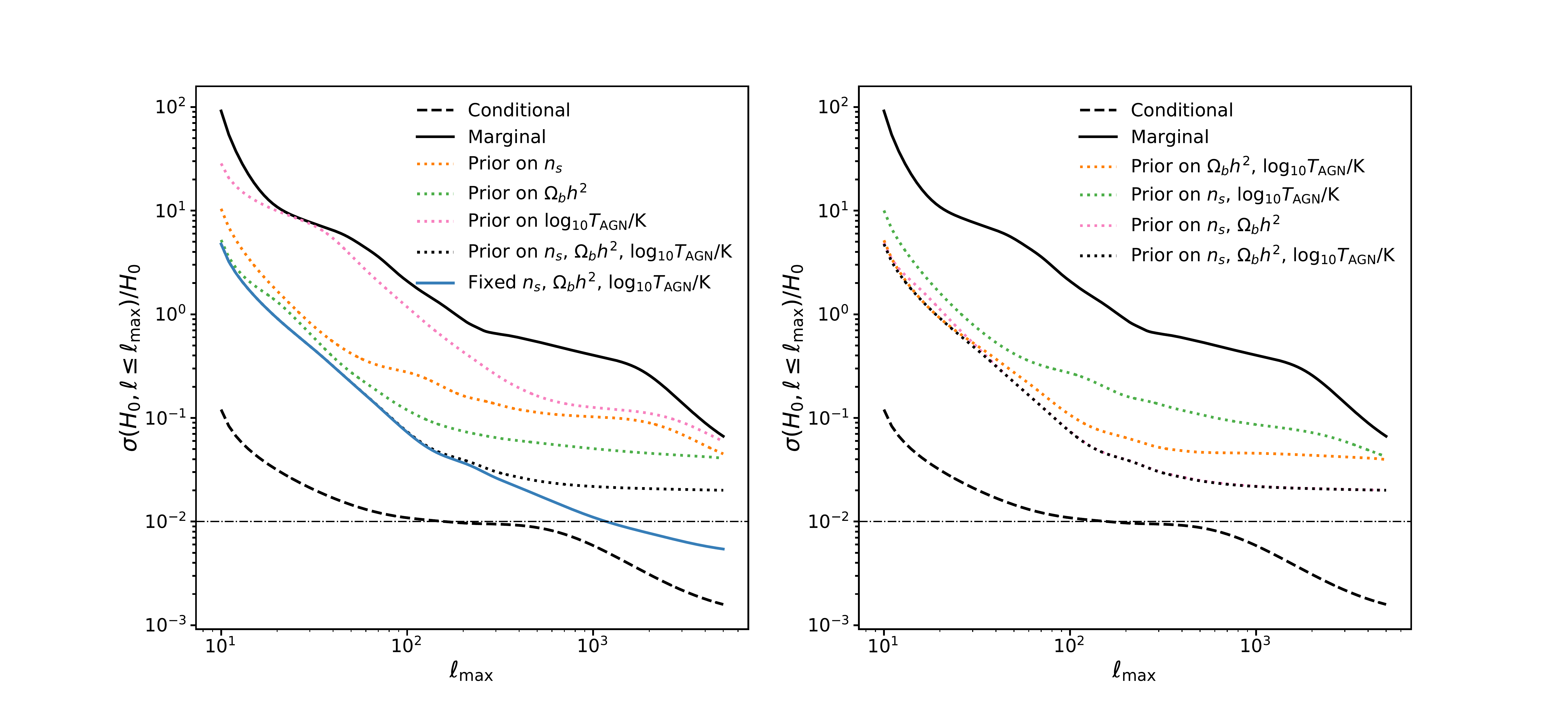}
  \includegraphics[width=\textwidth]{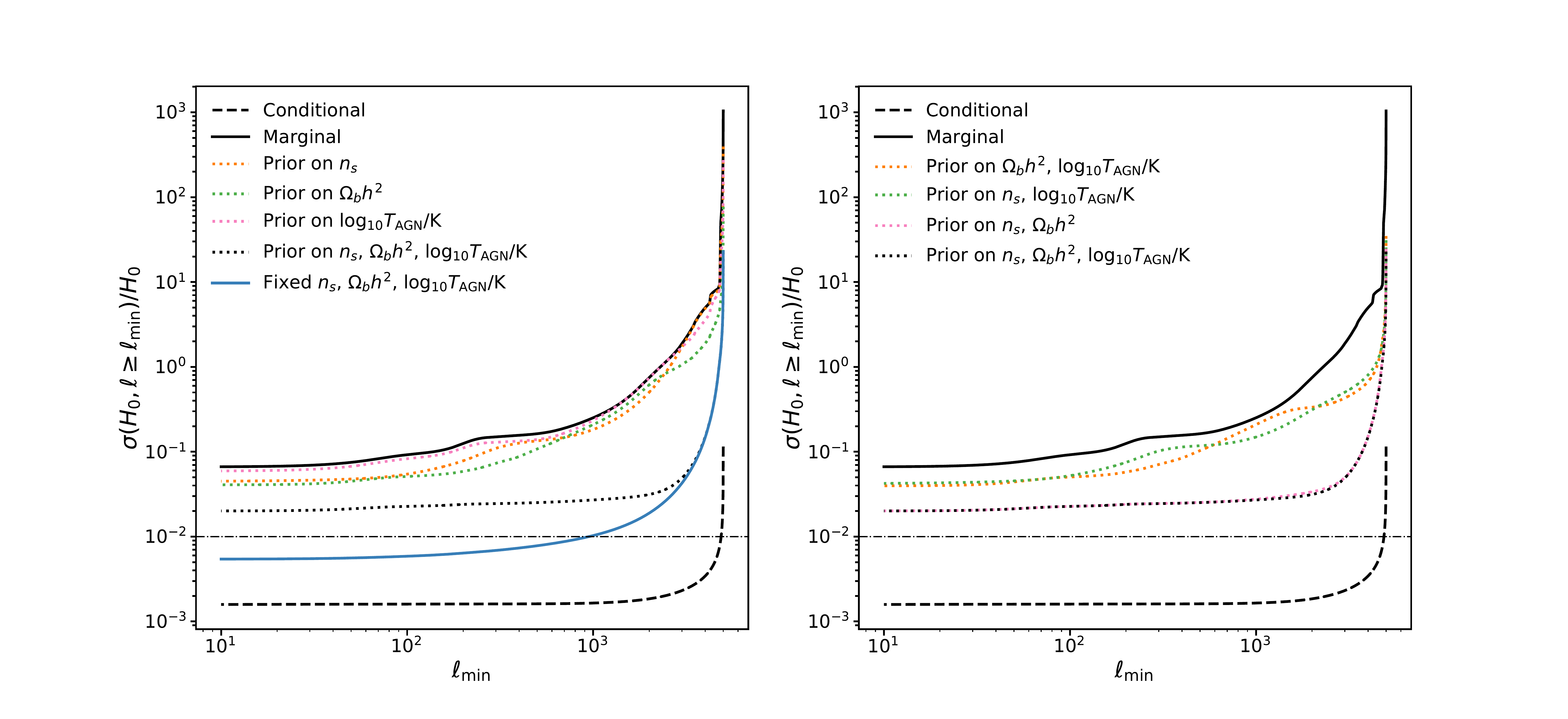}
  \caption{Same as Figure~\ref{fig:Euclid_H0_fixing} but imposing informative priors on $n_s$, $\Omega_b h^2$, and $\log_{10} T_{\mathrm{AGN}}/\mathrm{K}$ individually (left column) or in pairs (right column) and marginalising over the remaining parameters. The case where these three parameters are fixed is shown as the blue solid curve. In the top row we show constraints as a function of $\ell_{{\rm max}}$ fixing $\ell_{{\rm min}} = 10$, and in the bottom row we show constraints as a function of $\ell_{{\rm min}}$ fixing $\ell_{{\rm max}} = 5000$.}
  \label{fig:Euclid_H0_priors}
\end{figure*}

In Figure~\ref{fig:Euclid_H0_priors} we show forecast fractional constraints on $H_0$ for our toy Euclid-like survey as a function of both $\ell_{{\rm max}}$ and $\ell_{{\rm min}}$. This figure is the same as Figure~\ref{fig:Euclid_H0_fixing} except rather than fixing other parameters or groups of parameters we instead impose priors on those parameters. The priors we choose here are the `Planck lensing' priors given in Table~\ref{tab:priors} in the case of $n_s$ and $\Omega_bh^2$, and the simulation-informed prior on the baryon feedback parameter $\log_{10} T_{\mathrm{AGN}}/\mathrm{K}$ recommended by~\citet{2021MNRAS.502.1401M}.

The Figure demonstrates that the informative priors usually adopted in lensing-only analyses are not sufficient to give sub-percent constraints on $H_0$ in our toy survey. Further information is required, particularly on $n_s$, in order to break degeneracies between parameters that affect the slop of the lensing power spectrum in the same way as $H_0$.

If all three of $n_s$, $\Omega_b h^2$, and the baryon feedback amplitude are fixed, the bottom rows of Figure~\ref{fig:Euclid_H0_priors} shows that very large scales are not required to give sub-percent constraints on $H_0$. This suggests that the constraint is not coming from a measurement of the equality scale, but rather on the detailed shape of the power spectrum on non-linear scales. This suggests that the information may be diluted by correlations between measurements of the power spectrum at different $\ell$ coming from non-Gaussianity in the shear signal. It also hints at the potential for uncertainties in the baryon feedback modelling to affect the $H_0$ constraint.







\bsp	
\label{lastpage}
\end{document}